\documentclass[11pt]{article}
\usepackage{jheppub}
\usepackage{amsmath,amssymb}
\usepackage{calc}
\usepackage{jheppub} 

\usepackage{bm,amsmath,amssymb,slashed,graphicx,%
            enumerate,alltt,xspace,multirow,xcolor,mathrsfs}
\usepackage{fancyvrb}
\usepackage{booktabs}
\usepackage{graphicx}
\usepackage{subcaption}
\usepackage{xspace}
\usepackage[utf8]{inputenc}

\usepackage{url}
\usepackage{hyperref}

\usepackage{color}
\definecolor{darkgreen}{rgb}{0,0.5,0}
\definecolor{darkblue}{rgb}{0,0,0.7}
\definecolor{darkred}{rgb}{0.5,0,0.0}
\definecolor{darkorange}{rgb}{0.8,0.4,0.0}

\newcommand{\Ym}{Y$_{\text{m}}$}

\newcommand{\zcut}{z_{\text{cut}}}

\newcommand{\zetacut}{\zeta_{\text{cut}}}
\newcommand{\rhomin}{\rho_{\text{min}}}
\newcommand{\Mtop}{\mathcal{M}_{t \to bqq}}
\newcommand{\TopSplitter}{\texttt{TopSplitter}\xspace}

\makeatletter
\g@addto@macro\bfseries{\boldmath}
\makeatother

\title{Top tagging : an analytical perspective}

\author[a]{Mrinal Dasgupta,}
\author[b]{Marco Guzzi,}
\author[a]{Jacob Rawling,}
\author[c]{and Gregory Soyez}
\affiliation[a]{Lancaster-Manchester-Sheffield Consortium for Fundamental Physics, School of Physics
  \& Astronomy, University of Manchester, Manchester M13 9PL, United Kingdom}
\affiliation[b]{Department of Physics, Kennesaw State University,
  Kennesaw, GA 30144, USA}
\affiliation[c]{IPhT, CEA Saclay, CNRS UMR 3681, Universit\'e Paris-Saclay, F-91191 Gif-Sur-Yvette, France}

\emailAdd{mrinal.dasgupta@manchester.ac.uk}
\emailAdd{mguzzi@kennesaw.edu}
\emailAdd{jrawling@cern.ch}
\emailAdd{gregory.soyez@ipht.fr}


\abstract{In this paper we study aspects of top tagging from first
  principles of QCD.  We find that the method known as the CMS top
  tagger becomes collinear unsafe at high $p_t$ and propose variants
  thereof which are IRC safe, and hence suitable for analytical
  studies, while giving a comparable performance to the CMS tagger. We
  also develop new techniques to identify three-pronged jet
  substructure, based on adaptations of the Y-splitter method and its
  combination with grooming.  A novel feature of our studies, relative
  to previous calculations of two-pronged substructure, is our use of
  triple-collinear splitting functions, which owes to the presence of
  two mass scales of the same order, $m_t$ and $m_W$, in the signal
  jet.  We carry out leading logarithmic resummed calculations for the
  various top-taggers, for both background and signal jets, and
  compare the results to those from parton showers.  We also identify
  and comment on the main features driving tagger performance at high
  $p_t$ and discuss the role of non-perturbative effects. }
\begin{document}

\maketitle

\section{Introduction}\label{sec:intro}
Recent years have seen the field of jet substructure mature and
develop into one of the key areas of current LHC phenomenology
\cite{Seymour:1993mx, Butterworth:2002tt, Butterworth:2008iy,
  Ellis:2009me, Ellis:2009su, Krohn:2009th, Abdesselam:2010pt,
  Altheimer:2012mn, Altheimer:2013yza,Adams:2015hiv,Larkoski:2017jix}.
Amongst the numerous applications of substructure methods there are
direct searches for new physics beyond the standard model
\cite{Sirunyan:2018omb,Sirunyan:2018hsl,CMS:2018tuj,Aaboud:2018juj,
  Aaboud:2018zba, Aaboud:2018mjh}, crucial studies of the Higgs
sector of the standard model \cite{Sirunyan:2017dgc}, precise
determination of the top quark mass \cite{Hoang:2017kmk}, testing high
precision calculations for jets in QCD
\cite{Marzani:2017mva,Marzani:2017kqd,
  Frye:2016aiz,CMS:2017tdn,Aaboud:2017qwh}, and studies involving jets
in heavy-ion collisions \cite{Mehtar-Tani:2016aco,Connors:2018zpg}.
Following the commencement of the LHC run 2 at 13 TeV, electroweak
scale particles including the top quark can be extremely boosted,
which means their hadronic decays will often result in a single
jet. In such situations jet substructure studies have been proven to
provide important input which is the key to effectively distinguishing
signal from background as well as to improve resolution of signal mass
peaks.

One of the most active areas within the field of jet substructure has
been the study of boosted top quarks and several techniques are
available to study boosted tops including various ``top-taggers"
\cite{Kaplan:2008ie,
  Plehn:2009rk,CMS:2009lxa,CMS:2014fya,cmssw-top-tagger,Brooijmans:2008zza,Thaler:2008ju,Kasieczka:2015jma},
template tagging \cite{Almeida:2010pa}, shower deconstruction
\cite{Soper:2012pb}, jet shape variables such as N-subjettiness
\cite{Thaler:2010tr, Thaler:2011gf}, energy correlation functions
\cite{Larkoski:2013eya,Larkoski:2014zma,Moult:2016cvt} and
multi-variate methods exploiting machine learning
\cite{deOliveira:2015xxd, Baldi:2016fql,
  Barnard:2016qma,Komiske:2017aww,Kasieczka:2017nvn}.  The performance
of these tools has been investigated in detail using studies based on
Monte Carlo event generators.  Many of the above mentioned methods are
also increasingly used in experimental analyses at the LHC
\cite{Asquith:2018igt}.

An alternative approach to traditional Monte Carlo studies of jet
substructure has emerged and gained substantial ground in recent years
\cite{Dasgupta:2013ihk,
  Dasgupta:2013via,Larkoski:2013eya,Larkoski:2014wba,Larkoski:2015kga,Dasgupta:2015lxh,Dasgupta:2016ktv,Larkoski:2017cqq}. This
new approach is based on directly using perturbative QCD calculations
for jet substructure observables. Since the boosted regime with jet
masses $m \ll p_t$ is a classic multi-scale problem, and one
encounters the feature of large logarithms in $p_t/m$, perturbative
calculations at fixed-order in $\alpha_s$ are not directly useful on
their own, and one needs the techniques of analytic resummation to
give a satisfactory description of substructure observables in the
boosted limit .

Analytic resummed perturbative calculations have been shown to be
powerful methods in learning about jet substructure techniques often
yielding vital information about features that did not emerge in
shower studies prior to the advent of the analytics. Amongst some of
the benefits arising from analytical studies, one can list the
discovery of flaws such as kinks and bumps in the jet mass spectrum
with various taggers \cite{Dasgupta:2013ihk} which led to the
emergence of improved tools \cite{Dasgupta:2013ihk, Larkoski:2014wba},
the discovery of occasional issues with parton shower descriptions of
jet substructure \cite{Dasgupta:2013ihk}, the development of
observables which can be computed to high precision in QCD and which
display reduced sensitivity to non-perturbative effects
~\cite{Dasgupta:2013ihk, Larkoski:2014wba}, giving rise to
phenomenological studies with LHC data
\cite{CMS:2017tdn,Aaboud:2017qwh}. The analytical calculations give
powerful insight into the physics of jet substructure and into the
factors influencing and driving tagger performance in a way that is
virtually impossible to extract from limited shower studies unguided
by any analytics. There are several spin-offs arising from this
insight but most crucially it opens the way to creating optimal tools
which are not just performant but also reliable and robust.  It is, of
course, a relatively simple exercise to use Monte Carlo tools to get
an estimate of tagger performance, which is typically done via
generating the so-called ROC (Receiver Operating Characteristic)
curves which plot the background mistag rate against the signal
efficiency achieved with different taggers. However any result that
derives from QCD theory should also come with an uncertainty estimate
which reflects the theoretical approximations made and ROC curves are
no exception to this. However theoretical uncertainties on results
produced purely from Monte Carlo methods are not simple to estimate
and given the quite basic leading-logarithmic accuracy of parton
showers\footnote{Recent work has shown that some widely used parton
  showers often fail to achieve even full leading-logarithmic accuracy
  for well-known simple observables like the thrust distribution
  \cite{Dasgupta:2018nvj}.}  one may worry that such uncertainties, if
estimated properly, will be very large. Thus unambiguous statements
about comparative tagger performance based solely on Monte Carlo
studies are always potentially dangerous and support from analytic
calculations gains further importance.

In this article we shall carry out an investigation of aspects of top
tagging using analytic resummation as a main tool. As mentioned
before, detailed studies along these lines have already been
carried out for W/Z/H tagging and here our aim is to embark on a
similar level of understanding for top tagging.  We shall mainly
explore methods for identifying the top quark based on its
three-pronged decays i.e.\ shall focus on the prong finding aspect of
top taggers.

A study that covers all of the existing top-taggers goes beyond the
scope of our current article. Instead, to illuminate some of the main
features we shall consider a standard widely used method, the CMS
top-tagger \cite{CMS:2009lxa,CMS:2014fya} which is closely related to
the Johns Hopkins tagger \cite{Kaplan:2008ie}, as well as introduce
another method based on the \Ym-splitter tagger
\cite{Dasgupta:2016ktv}, itself a variant of the Y-splitter method
already used in top-tagging
\cite{Butterworth:2002tt,Brooijmans:2008zza}, and also investigate the
combination of \Ym-splitter with jet grooming.  We start in section
\ref{sec:tagger-defs} by defining in detail the CMS tagger and
pointing out that it suffers from the issue that it becomes collinear
unsafe at high $p_t$. The collinear unsafety, apart from making the
tagger unreliable in general, rules out the possibility for any
all-orders resummed calculations for substructure observables defined
with the CMS tagger.  We therefore propose two new variants of the CMS
tagger, $\mathrm{CMS^{3p,mass}}$ and \TopSplitter, which are both
infrared and collinear (IRC) safe and, especially in the case of
\TopSplitter, more suited to analytical calculations based on
resummation.  Next, in the same section, we also define the
\Ym-splitter method and extend it for the purpose of identifying
three-pronged jet substructure in the context of top tagging. We
further discuss the combination of \Ym-splitter with grooming which is
needed in order to achieve a good performance with \Ym-splitter.

In section \ref{sec:analytic} we carry out an
$\mathcal{O} \left(\alpha_s^2\right)$ leading-order calculation for
the CMS tagger and for \Ym-splitter. At this order the CMS tagger is
IRC safe and the collinear unsafety arises at next-to--leading order
level and beyond. We first carry out a calculation using a simplified
picture based on soft emissions which are strongly ordered in emission
angles. Next we discuss reasons for why such a picture, which we might
expect to be correctly described by most parton shower methods, may be
insufficient for the case of top tagging. We explain that a more
natural picture to describe top-taggers is instead based on the use of
triple-collinear splitting functions which describe the collinear
$1 \to 3$ splitting of an energetic parton, with no strong ordering
between the final emissions and no soft approximation
\cite{Campbell:1997hg,Catani:1998nv,Catani:1999ss}. We then carry out
calculations for the various taggers using the triple-collinear
splitting functions and phase-space.

Section \ref{sec:analytic-resum} contains a description of the
resummation we perform for the different taggers starting with
\Ym-splitter and its combination with grooming and then moving on to
\TopSplitter. Here we present the arguments leading to the resummed
results in each case as well as leading-logarithmic results for the
Sudakov form factors using a fixed-coupling approximation,
although our final results also include both the effect of
hard-collinear next-to--leading logarithmic corrections as well as
running coupling effects. We discuss how to match the Sudakov form
factors computed in the soft and strongly ordered approximation, with
the leading-order pre-factor computed in the triple-collinear limit.

In section \ref{sec:results} we first discuss the numerical impact of
including the triple-collinear splitting function and of various
resummation effects, then compare the results of our analytical
calculations for QCD background jets with parton level results from
using the Pythia shower~\cite{Sjostrand:2014zea}. We study different
analytical approximations to the Sudakov exponent for each tagger
compared to the Pythia result and also directly compare the taggers to
one another both using our analytical results and using Pythia.

Section \ref{sec:signalandroc} contains our studies for signal jets as
well as studies of tagger performances with ROC curves generated both
analytically and with Monte Carlo. We also investigate in this section
the role of non-perturbative effects including both hadronisation and
the underlying event. Our conclusions are presented in section
\ref{sec:conclusions}. An explicit demonstration of the collinear
unsafety of the CMS tagger using fixed-order perturbative QCD, a
discussion of further tagger variants, and analytical results
including running-coupling effects can be found in the appendices.

\section{Tagger  definitions}
\label{sec:tagger-defs}

In this section we shall describe the default version of the CMS
tagger and discuss its potential collinear unsafety issue. We shall
define a variation of the CMS method,
$\mathrm{CMS^{3p,\mathrm{mass}}}$, that is IRC safe and we shall
introduce a new method we call \TopSplitter that apart from being IRC
safe is more amenable to a detailed analytical understanding. We also
discuss our implementation of the $\mathrm{Y_m}$-splitter method for
top tagging and discuss the combination of $\mathrm{Y_m}$-splitter
with grooming, extending the ideas we first introduced in
Ref.~\cite{Dasgupta:2016ktv}.

\subsection{The CMS top tagger and new methods}\label{sec:cms-definition}

The steps involved in the CMS top tagger are detailed below.  The
first version of the CMS tagger reported
in~\cite{CMS:2009lxa,CMS:2014fya}, proceeds as follows\footnote{The
  explicit code can be found as part of CMSSW, see
  \cite{cmssw-top-tagger} which is what we have used in this paper.}:

\begin{enumerate}
\item The initial anti-$k_t$ jet \cite{Cacciari:2008gp} is
  re-clustered using the Cambridge-Aachen algorithm
  \cite{Dokshitzer:1997in,Wobisch:1998wt}.

\item {\em Primary decomposition}: the last step of the clustering is
  undone, giving 2 prongs. These two prongs are examined for the
  condition
  \begin{equation}\label{eq:zetacut}
    p_t^{\mathrm{prong}} > \zeta_{\text{cut}} \, p_t^{\text{jet}},
  \end{equation}
  where $p_t^{\text{jet}}$ refers to the hard jet transverse
  momentum. $\zeta_{\text{cut}}$, referred to as $\delta_P$ in the CMS
  papers, is a parameter which is usually taken as $0.05$.
  If both prongs pass the cut then the ``primary'' decomposition
  succeeds.  If both prongs fail the cut then the jet is rejected
  i.e.\ is not tagged as a top jet.  If a single prong passes the cut
  the primary decomposition recurses into the passed prong, until the
  decomposition succeeds or the whole jet is rejected.
  Note that during the recurrence, $p_t^{\text{jet}}$ (used in
  (\ref{eq:zetacut})) is kept as the transverse momentum of the
  original jet.
  
\item {\em Secondary decomposition}: with the two prongs found by the
  primary decomposition, repeat the declustering procedure as for
  the primary decomposition, still defining the $\zeta_{\text{cut}}$
  condition~\eqref{eq:zetacut} wrt the original jet $p_t$.
  This can result in either both prongs from the primary decomposition
  being declustered into two sub-prongs, only one prong being
  declustered, or none.
  When no further substructure is found in a primary prong, the
  primary prong is kept intact in the final list of prongs. When two
  sub-prongs are found both are kept in the final list of prongs.
  Ultimately, this leads to two, three or four prongs emerging from
  the original jet. Only jets with three or four sub-prongs are then
  considered as top candidates.
  
\item Taking the three highest $p_t$ subjets (i.e.\ prongs) obtained by
  the declustering, the algorithm finds the minimum pairwise mass and
  requires this to be related to the $W$ mass, $m_W$, by imposing the
  condition
  $\mathrm{min} \left(m_{12},m_{13},m_{23} \right) > m_{\mathrm{min}}
  $ with $m_{\mathrm{min}} \lesssim m_W$.
  For practical applications, $m_{\text{min}}$ is usually taken as
  $50$~GeV.

\item Note that in the second version of the
  tagger~\cite{CMS:2014fya}, the decomposition procedure also imposes
  an angular cut: when examining the decomposition of a subjet $S$
  into two prongs $i$ and $j$, the CMS tagger also requires
  $\Delta R_{ij} > 0.4 - A p_t^S$ where
  $\Delta R_{ij} = \sqrt{\Delta y_{ij}^2+ \Delta \phi_{ij}^2}$ and
  $p_t^S$ refers to the transverse momentum of the
  subjet.\footnote{For the $p_t$ scale entering the $\Delta R$
    condition, Ref.~\cite{CMS:2014fya} mentions using the original jet
    (resp. the primary prongs) during the primary (resp. secondary)
    decomposition. However, the code in CMSSW is explicitly using the
    ``local'' subjet $p_t$.}
  The default value for $A$ is $0.0004 \, \mathrm{GeV^{-1}}$.
\end{enumerate}

We also note here that the first version of the tagger
\cite{CMS:2009lxa} does not make a reference to the $\Delta R$
condition in the decomposition of a cluster. In fact without a
$\Delta R$ cut the tagger is \emph{collinear unsafe}. This in turn
implies that fixed-order perturbative QCD results for observables can
produce divergent results, thereby compromising the reliability of the
tagger.

The collinear unsafety arises due to the process of selecting the
three hardest prongs out of four prongs (to define the
$m_{\mathrm{min}}$ cut) which is sensitive to arbitrarily collinear
hard radiation (see Appendix~\ref{app:event2} for an explicit
demonstration of the collinear unsafety aspect, using fixed-order
perturbative QCD).
With a $\Delta R$ cut formal collinear safety is restored but for
small values $\Delta R \ll 1$, one will encounter large logarithms in
$\Delta R$ making a perturbative description of the tagger potentially
complicated. Also, given the recommended optimal value for the
parameter $A$, as one progresses towards high $p_t$ values the
$\Delta R$ cut becomes smaller and eventually vanishes which means
that the default CMS tagger will again be collinear unsafe at
asymptotically large $p_t$. 

To evade the issue of collinear unsafety one could argue that
precision perturbative calculations are not the main aim of jet
substructure studies, at least in the context of LHC searches for new
physics.  However as we stated in the introduction, assessing the
uncertainty on results for tagger signal and background efficiencies
is far from simple, and with an IRC unsafe tool this becomes
impossible. Hence any statements about tagger performance based on ROC
curves cannot be formally taken at face value.  Moreover not all jet
substructure studies are aimed at direct searches for new physics, and
substructure tools are widely used in an increasing variety of
contexts including for precision studies and comparison between
perturbative QCD calculations and experimental data
\cite{Marzani:2017mva, Marzani:2017kqd,Frye:2016aiz}, possible
extractions of the strong coupling \cite{Bendavid:2018nar}, and in the
case of top quark physics, determinations of the top mass
\cite{Hoang:2017kmk}. For such studies, where high precision and small
uncertainties are essential, any IRC unsafety issues can severely
compromise the validity of the results obtained and conclusions
reached.  It is therefore desirable to ensure a set of substructure
tools that are free from IRC unsafety issues while still yielding the
required performance.

Ultimately, this collinear unsafety issue motivated us to investigate
alternatives to the $\Delta R$ cut imposed by the CMS top tagger and
to introduce the following new methods:

\begin{itemize}
\item {$\mathrm{CMS^{3p,\mathrm{mass}}}$}: say that the primary
  decomposition led to the two prongs $\mathrm{A}$ and $\mathrm{B}$
  and that prong $\mathrm{A}$ has a secondary decomposition into
  subprongs $\mathrm{A^\prime}$ and $\mathrm{A^{\prime \prime}}$ while
  $\mathrm{B}$ is decomposed into $\mathrm{B^\prime}$ and
  $\mathrm{B^{\prime \prime}}$. Rather than selecting the hardest 3
  objects from the set ${\mathrm{A^\prime}},$
  $\mathrm{A^{\prime \prime}}$, $\mathrm{B^\prime}$,
  $\mathrm{B^{\prime \prime}}$ as in the standard CMS tagger, one
  instead examines the invariant masses
  $m^2_{\mathrm{A^{\prime}}\mathrm{A^{\prime
        \prime}}}=(p_{\mathrm{A^{\prime}}}+p_{\mathrm{A^{\prime
        \prime}}})^2$
  and
  $m^2_{\mathrm{B^{\prime}}\mathrm{B^{\prime
        \prime}}}=(p_{\mathrm{B^{\prime}}}+p_{\mathrm{B^{\prime
        \prime}}})^2$.
  If
  $m^2_{\mathrm{A^\prime A^{\prime \prime}}} > m^2_{\mathrm{B^{\prime}
      B^{\prime \prime}}}$
  then one simply considers the 3 prongs to be $\mathrm{A^{\prime}}$,
  $\mathrm{A^{\prime\prime}}$ and $\mathrm{B}$, and vice-versa. In
  this variant of the CMS method we obtain 3 prongs which can be used
  in the $m_{\mathrm{min}}$ condition without any collinear unsafety
  issues and without a $\Delta R$ cut. We shall refer to this variant
  as $\mathrm{CMS^{3p,\mathrm{mass}}}$ since it produces three prongs
  based on a selection using invariant masses.
  
\item{\TopSplitter}: As we shall clarify in more detail in subsequent
  sections, it proves to be advantageous in some respects to nominate
  the emission that would dominate the mass of a prong in the limit
  where all emissions are soft and strongly ordered in mass, as a
  product of the declustering, instead of the largest-angle emission
  passing the $\zetacut$ as given by the C-A declustering.
  In order to do so we first keep the same procedure as above for
  identifying the two prongs A and B that emerge from the primary
  decomposition. Now consider the decomposition of each of these
  prongs starting say with prong A. We decluster this precisely as
  before until we find an emission $i$ that passes the $\zetacut$
  condition.  At this stage however we also consider {\emph{all}}
  subsequent emissions further down the C-A tree following the hardest
  branch, together with emission $i$, and identify the emission $j$ in
  this set that has the largest value of $p_{tj} \theta_j^2$, i.e
  contributes the most to the prong mass in the limit that all
  emissions are soft.\footnote{Given that this emission is either
    emission $i$ itself or a smaller angle emission, it is clear that
    it must also pass the $\zetacut$ condition.}  We take this
  emission to be $\mathrm{A}^{\prime \prime}$ i.e.\ one of the products
  of the declustering of A.  The other product of the declustering is
  labelled $\mathrm{A^{\prime}}$ as before. It consists of the
  remaining object to which $\mathrm{A}^{\prime \prime}$ is clustered
  in the C-A clustering sequence, along with all emissions preceding
  $\mathrm{A}^{\prime \prime}$ in the C-A tree which passed the
  $\zetacut$ condition, such as emission $i$. We call this new method
  \TopSplitter.

\end{itemize}

Other variants are possible and they will be discussed in
Appendix~\ref{app:variants}.

\subsection{The \Ym-splitter method for top tagging}\label{sec:Ymsplittertagger-def}

The use of the Y-splitter method for top tagging was already
considered by Brooijmans and made use of in ATLAS studies of top
tagging~\cite{Brooijmans:2008zza,TheATLAScollaboration:2015bkc}.

In Refs.~\cite{Dasgupta:2015yua,Dasgupta:2016ktv} it was found that
the Y-splitter technique, when supplemented with grooming, was a
high-performance method for the tagging of electroweak scale particles
that exhibit two pronged decays, especially for $p_t$ values in the
TeV range. To be more precise, it was observed in
Refs.~\cite{Dasgupta:2015yua,Dasgupta:2016ktv} that Y-splitter gives
an excellent suppression of QCD background jets due to a large Sudakov
suppression factor. However the performance of Y-splitter on signal
jets was poor as the lack of an explicit grooming step resulted in
loss of signal. Once grooming is performed after Y-splitter (either
via mMDT \cite{Dasgupta:2013ihk} or trimming \cite{Krohn:2009th}),
while the feature of the background suppression stays largely intact,
there is considerable improvement in the signal efficiency. This
results in striking gains for the signal significance.

Therefore it also becomes of interest to adapt Y-splitter with
grooming to the case of top decays. In Ref.~\cite{Dasgupta:2016ktv} we
introduced and discussed several variants of the Y-splitter technique
for the case of two pronged decays. The variant that emerged as both
most robust and performant was a variant we called \Ym-splitter, which
makes use of the gen-$k_t$~($p=1/2$) algorithm~\cite{Cacciari:2011ma}
to define distances between objects, in place of the $k_t$ distance
\cite{Catani:1991hj,Catani:1993hr,Ellis:1993tq} used in the standard
Y-splitter.  The use of the gen-$k_t$~($p=1/2$) distance (hereafter
referred to just as the gen-$k_t$ distance for brevity) guarantees an
ordering equivalent to an ordering in mass in the soft limit which
facilitates the direct analytical understanding of the tagger
behaviour, with the fringe benefit of giving a slightly better
performance compared to the standard $Y$-splitter.

We will also consider pre-grooming with SoftDrop ($\beta=2$ and
$\beta=0$ (i.e.\ mMDT)) prior to the application of
\Ym-splitter. The $\beta=2$ pre-grooming option was already explored
for the tagging of W/Z/H and found to give good performance while
highly reducing the sensitivity to non-perturbative effects
\cite{Dasgupta:2016ktv}. The $\beta=0$ pre-grooming option was not
considered in Ref.~\cite{Dasgupta:2016ktv} since for the case of W/Z/H
tagging studied there, this option was found to reduce the important
Sudakov suppression of the background. In the present case however,
where we have a coloured object being tagged, the situation will be
different as we shall explain in more detail in
section~\ref{sec:signalandroc}, and pre-grooming with mMDT becomes a
useful option to consider.

To adapt \Ym-splitter for use in top tagging one considers applying it
twice in succession, as follows:

\begin{enumerate}
\item Perform a primary decomposition of the initial fat
  jet by doing a first declustering but here based on the
  gen-$k_t$~($p=1/2$) distance measure. On each of the two
  prongs obtained by undoing the clustering apply the
  $\zeta_{\text{cut}}$ condition, Eq.~(\ref{eq:zetacut}). If the
  $\zeta_{\text{cut}}$ condition fails for either of the two prongs,
  discard the jet as a top candidate, otherwise move to the next step.
\item Decluster both prongs obtained from the primary decomposition
  (still using the gen-$k_t$ algorithm). The prong that produces the
  smaller gen-$k_t$ distance in the declustering is kept
  unaltered. The prong that yielded the larger gen-$k_t$ distance is
  tested for the $\zeta_{\text{cut}}$ condition as for the primary
  decomposition. If the $\zeta_{\text{cut}}$ condition passes proceed
  to the next step otherwise the jet is rejected.
\item Take the three prongs that emerge after the secondary
  decomposition (i.e.\ the unaltered primary prong and the two
  secondary prongs which passed the $\zetacut$ condition) and impose
  the $m_{\mathrm{min}}$ condition on the minimum pairwise mass.
\end{enumerate}

Additionally, as mentioned above, we shall consider pre-grooming with
mMDT and SoftDrop on the full jet, prior to the application of the
above steps. Lastly, we also introduced additional variants for
\Ym-splitter similar to the case of the CMS tagger and these are also
discussed in Appendix~\ref{app:variants}.

\section{Analytical calculations at fixed-order}\label{sec:analytic}

In this section we shall carry out some basic leading-order analytic
calculations to help us better understand the action of top taggers on
QCD jets. We shall start by using a soft and collinear approximation
for emissions within the jet and then discuss improving this
approximation in light of the specific requirements for three-pronged
jet substructure and top taggers.

\subsection{Leading-order calculations in the soft-collinear limit}\label{sec:analytic-leading}

The standard idea that is exploited in two-body tagging to distinguish
signal from background is to exploit the differences in splitting
functions between QCD decays and those involving W/Z/H. While the
former contain soft enhancements, the latter are regular in the soft
limit and hence cutting the soft region via a $\delta_P$ or
$z_{\mathrm{cut}}$ type of condition\footnote{This would involve a
  cut of the form
  $\frac{\mathrm{min}\left(p_{T,i},p_{T,j}\right)}{p_{T,i}+p_{T,j}}
  >z_{\mathrm{cut}}$
  which uses the local $p_T$ of the cluster being decomposed,
  i.e.\ $p_{T,i}+p_{T,j}$ instead of the global $p_T$ of the hard jet
  in the denominator as is the case for the original CMS $\delta_P$
  condition.} reduces the background significantly compared to the
modest impact on the signal (see e.g.\ Ref.~\cite{Dasgupta:2013ihk} for
explicit examples and more details). For the case of three-body
hadronic top decays we have instead {\em two} branchings that are not
soft-enhanced namely the branching $t \to b \,W$ and then the two-body
W decay to quarks. We should therefore expect that the double
application of the $\zetacut$ condition exploits this feature.

In order to see this most clearly, in this sub-section we perform a
leading-order QCD calculation for the jet-mass distribution for QCD
jets after the application of top-tagging methods. In the boosted
limit the jet mass $m$ is small compared to the jet $p_t$ and we shall
work in terms of the standard variable, invariant under boosts along
the jet direction, $\rho = \frac{m^2}{R^2 p_t^2}$, with $m$ the jet
mass and $R$ the jet radius, so that $\rho \ll 1$. For the application
of top taggers, aside from the jet mass we also have the
$m_{\mathrm{min}}$ condition and hence also define
$\rho_{\mathrm{min}}= \frac{m_{\mathrm{min}}^2}{R^2 p_t^2} \ll 1.$ The
other parameter which enters our calculations is $\zetacut$. This is
chosen not too small in order to reduce the QCD background
i.e.\ $\zetacut \gg \rho, \rho_{\mathrm{min}}$ but nevertheless
$\zetacut \ll 1$, with the value $\zetacut =0.05$ generally favoured
in practical applications. We therefore expect that in a perturbative
calculation we will encounter large logarithms in the jet masses
$\rho, \rho_{\mathrm{min}}$ as well as large logarithms in $\zetacut$
but with the former being numerically dominant over the latter.

A further issue that arises is the potential presence of logarithms of
$\rho/\rho_{\mathrm{min}}$ at each order in perturbation theory. For
simplifying the leading-order calculations in this subsection and in
order to most conveniently illustrate the role of the top-taggers we
shall assume that $\frac{\rho}{\rho_{\mathrm{min}}} \gg \zetacut$ so
that logarithms of $\zetacut$ may be neglected compared to those in
$\rho/\rho_{\mathrm{min}}$. In practice however, given that we are
interested in top tagging, the jet mass $m \sim m_t$ and
$m_{\mathrm{min}} \sim m_W$ are not strongly ordered, hence logarithms
of $\rho/\rho_{\mathrm{min}}$ are not necessarily large and cannot
generally be taken to be dominant over logarithms of $\zetacut$ . We
shall return to address these points in the next subsection and
subsequent sections.

With the above mentioned large logarithms in mind we shall initially
specialise to the soft and collinear limit for all emissions
i.e.\ $z_i, \theta_i \ll 1$, where $z_i$ is the fraction of the jet's
$p_t$ carried by emission $i$ and $\theta_i$ the angle of emission $i$
wrt the jet axis.  Moreover to calculate the leading logarithms in jet
mass we can further assume that successive emissions are strongly
ordered in angles.  In order to pass the top-tagger conditions one
requires at least two emissions in addition to the hard parton that
initiates the jet. Thus the leading order in perturbative QCD for the
jet mass distribution, with application of top tagging, is order
$\alpha_s^2$. Assuming that the jet is initiated by a hard quark we
start by considering two soft and collinear gluon emissions strongly
ordered in emission angles and emitted independently by the hard
quark, corresponding to a $C_F^2$ colour factor.

We start by applying the CMS top tagger and variants thereof. At the
leading order, i.e.\ order $\alpha_s^2$, the CMS tagger,
$\mathrm{CMS^{3p,\mathrm{mass}}}$ and \TopSplitter are all
equivalent. The IRC unsafety issue of the CMS tagger occurs at order
$\alpha_s^3$ i.e.\ at the NLO level in the context of the present
calculations. Hence for the purpose of this section we shall refer
explicitly to the CMS tagger with the understanding that the results
apply equally for our new methods.
After the primary C-A  declustering of the jet, the
larger angle gluon $k_1$ emerges first and is subjected to the
$\zetacut$ condition which leads to the constraint $z_1 >\zetacut$.
We obtain two subjets: a massless subjet $j_1$ consisting of parton
$k_1$ and a massive subjet $j_2$ composed of a hard quark with
four-momentum $p$ and the emission $k_2$.  One then declusters $j_2$
into its massless partonic constituents and retains the jet only if
$z_2 >\zetacut$. 

At the leading-logarithmic level we can assume that the jet mass
$\rho$ is dominated by the contribution from the larger-angle emission
$k_1$. This can be shown to be correct up to terms involving only
subleading logarithms in $\rho/\rho_{\mathrm{min}}$, albeit enhanced
by logarithms of $\zetacut$. Hence we have that
$\rho \sim z_1 \theta_1^2$ where all angles are taken to be measured
in units of the jet radius $R$, meaning in particular that they should
be less than 1.

The tagger then places a constraint on the minimum pairwise mass of
the three partons $p$, $k_1$ and $k_2$ which can be written as 
$\mathrm{min} \left(z_1 \theta_1^2, z_2 \theta_2^2, z_1 z_2
  \theta_{12}^2\right) > \rhomin$.
Since at leading-logarithmic accuracy $z_1 \theta_1^2$ dominates the
jet mass, the minimum pairwise mass is the minimum of $z_2 \theta_2^2$
and $z_1 z_2 \theta_{12}^2$.
In the strongly ordered limit we have that $\theta_1 \gg \theta_2$ and
$\theta_{12} \approx \theta_1$.
Therefore, the minimum pairwise mass can be taken to be $z_2
\theta_2^{2}$, up to subleading $\ln \zetacut$ corrections once again.

Using $\theta_{12} \approx \theta_1$, one can carry out the leading
logarithmic calculation straightforwardly. The jet mass distribution
is approximated as follows:
\begin{multline}
\label{eq:LLbasic}
\frac{1}{\sigma}\left(\frac{d\sigma}{d\rho}\right)^{\mathrm{LO, soft-collinear}} =  \bar{\alpha}^2 \int \frac{dz_1}{z_1}
\frac{dz_2}{z_2} \frac{d\theta_1^2}{\theta_1^2}
\frac{d\theta_2^2}{\theta_2^2}   \times\Theta(\theta_2^2<\theta_1^2<1)\, \delta(\rho-z_1
  \theta_1^2)  \\
\Theta(z_1>\zetacut)\,\Theta(z_2>\zetacut)\, \Theta( z_2 \theta_2^2 > \rho_{\mathrm{min}} ),
\end{multline}
where we defined $\bar{\alpha} = \frac{C_F \alpha_s}{\pi}$, taking for
definiteness the case of a quark initiated jet.

Recalling that we assumed, purely for the sake of simplicity, a
hierarchy of masses such that $\rhomin /\rho \ll \zetacut$ then one
obtains the simple result

\begin{equation}
\label{eq:cmsll}
\frac{1}{\sigma}\left(\frac{d\sigma}{d\rho}\right)^{\mathrm{LO,soft-collinear}}= \frac{\bar{\alpha}^2}{\rho} \left(\ln^2
\frac{1}{\zetacut} \ln \frac{\rho}{\rhomin} + \mathcal{O}
\left( \ln^3 \zetacut \right) \right).
\end{equation}

The essential functioning of the tagger at leading-order is encoded in
the above equation. For comparison, remember that the leading
logarithmic behaviour for the QCD background jet mass distribution is
double logarithmic i.e.\ $\frac{\rho}{\sigma} \frac{d\sigma}{d\rho} \sim \bar{\alpha}^2
\ln^3\frac{1}{\rho}$.  After applying the CMS tagger two large
logarithms in jet mass have been replaced by logarithms of $\zetacut$
which are not essentially large. This is similar to the action of
taggers in the two-body case. The result now contains only a
potentially large logarithm in $\rho/\rho_{\mathrm{min}}$ coming from
the $m_{\text{min}}$ condition.

We can also perform a similar calculation for the \Ym-splitter
technique defined in section~\ref{sec:Ymsplittertagger-def}. The
essential difference with the CMS tagger is the use of the gen-$k_t$
distance with its parameter $p$ taken to be $1/2$ (instead of the C-A
declustering used for the CMS tagger). The distances to be considered
are then $z_1\theta_1^2, z_2 \theta_2^2$ and
$\mathrm{min} (z_1,z_2) \theta_{12}^2$. As for the pairwise mass
constraint for the CMS tagger considered above, one always obtains in
our approximation that $z_2 \theta_2^{2}$ is the smallest distance, up
to subleading corrections.

Thus on declustering the jet, emission $k_1$ emerges first (and is
required to satisfy the $\zeta_{\mathrm{cut}}$ condition) along with a
massive prong consisting of the hard initiating quark and the emission
$k_2$. A secondary declustering of the massive prong then yields
emission $k_2$ which is also required to pass the
$\zeta_{\mathrm{cut}}$ condition.  Also, as before, the minimum
pairwise mass of the three partons is given by $z_2 \theta_2 ^2$.
Therefore the leading-order result we obtain for \Ym-splitter is the
same as that for the CMS tagger reported in Eq.~\eqref{eq:cmsll}.

Similar calculations can be carried out for the terms involving
secondary emissions i.e those involving say an initial quark emitting
a gluon which splits to a $gg$ or $q \bar{q}$ pair i.e.\ the $C_F C_A$
and $C_F T_R n_f$ channels. In the former case one obtains again a
logarithm in $\rho/\rho_{\mathrm{min}} $ also with an accompanying
$\ln^2 \zetacut$ coefficient while the $C_F n_f$ term has also the
logarithm in $\rho/\rho_{\mathrm{min}} $ but with only a
$\ln \zetacut$ coefficient due to the absence of a soft singularity in
the $g \to q\bar{q}$ splitting.

While Eq.~\eqref{eq:cmsll} captures the basic physics of the tagger in
the limit $1 \gg \rho \gg \rho_{\mathrm{min}}$, a number of comments
are in order. First of all we have used the approximation of strong
angular-ordering which is intended to capture logarithms in
$\rho/\rho_{\mathrm{min}}$. Additionally we also used the soft
approximation in performing the calculation which is sufficient to
generate the logarithms of $\zetacut$ reported in
Eq.~\eqref{eq:cmsll}, but not constant terms stemming from hard
collinear emissions or terms involving powers of $\zetacut$.  The
former constant contributions, in particular, are known to be
numerically significant in practice~\cite{Dasgupta:2013ihk}. The
standard method to include hard collinear splitting is to correct the
soft approximation, used above, with the full splitting function
i.e.\ make the replacement $\frac{ dz}{z} \to \frac{1+(1-z)^2}{2z} dz$
for the integral over energy fractions in Eq.~\eqref{eq:LLbasic}. We
should then expect a product of leading-order splitting functions to
appear, which account for both hard branchings i.e.\ the region where
$z_1,z_2$ are both finite.
Moreover, beyond the soft limit, the gen-$k_t$ distances, involved in
the \Ym-splitter calculation, would no longer be identical to the
mass. All these changes are straightforward to implement and do not
require a fundamental change of the basic angular-ordered picture
above.

More crucially perhaps, as we already observed, the approximation
$\rho \gg \rho_{\mathrm{min}}$, while convenient analytically, is in
practice not a good approximation for the case of top tagging. Without
any strong ordering between $\rho$ and $\rho_{\mathrm{min}}$ we are
led to a situation where the only genuinely large logarithms in the
boosted limit are those in $\rho$ or equivalently
$\rho_{\mathrm{min}}$ but not those of $\rho/\rho_{\mathrm{min}}$.  In
other words we should regard Eq.~\eqref{eq:cmsll} as an approximation
to a result of the form
\begin{equation}
\label{eq:cmsll2}
\frac{1}{\sigma}\left(\frac{d\sigma}{d\rho} \right)^{\mathrm{LO,\mathrm{triple-collinear}}}= \frac{\alpha_s^2}{\rho} f_q \left
  (\rho,\rho_{\mathrm{{min}}}, \zetacut\right).
\end{equation}
In the above equation $f_q$ is a function that needs to be computed in
full i.e.\ without any soft or collinear approximation and where the
suffix $q$ indicates a quark initiated jet. It contains the
contributions from $C_F^2, C_F C_A$ and $C_F T_R n_f$ colour factors
on an equal footing.  The only approximation inherent in writing
Eq.~\eqref{eq:cmsll2} is the approximation of small $\rho \ll 1$,
corresponding to appearance of the $1/\rho$ factor, which is justified
by working in the boosted limit $m^2 \ll p_t^2$.  Thus we need to
examine the collinear decay of an initial parton to three partons,
producing a small jet mass $\rho$, but with {\emph{no ordering between
    the three partons themselves, in either energy or angle}}. The
appropriate extension of Eq.~\eqref{eq:cmsll} requires the use of
\emph{triple-collinear $(1 \to 3)$ splitting functions}. Calculations
based on these shall be the subject of the next section.

\subsection{The triple-collinear limit of a QCD jet}\label{sec:analytic-triple}

Here we shall use the $1 \to 3$ splitting
functions~\cite{Campbell:1997hg,Catani:1998nv,Catani:1999ss} to compute the
differential distribution in the jet mass $\rho$, for the CMS and
\Ym-splitter methods.

Consider for example the collinear decay of an initial quark to a
quark and two gluons, taking the Abelian $C_F^2$ term of the
triple-collinear splitting functions as an example. The explicit
functional form for the spin-averaged splitting function is
\begin{equation}
\label{eq:CF2}
\langle \hat{P} ^{(\mathrm{ab})}_{g_1 g_2 q_3}\rangle = C_F^2 \left [ \frac{s^2_{123}}{2s_{13} s_{23}}  z_3
\left(\frac{1+z_3^2}{z_1 z_2}  \right)+\frac{s_{123}}{s_{13}} \left(
  \frac{z_3(1-z_1)+(1-z_2)^3}{z_1 z_2} \right)-\frac{s_{23}}{s_{13}}
\right ]
+\left(1 \leftrightarrow 2 \right).
\end{equation}
For the other colour configurations (involving $C_A$ and $n_f$), we
refer the reader to the original references \cite{Campbell:1997hg,
  Catani:1998nv,Catani:1999ss}. Here $s_{ij}$ and $s_{ijk}$ are the
usual kinematic invariants $(p_i+p_j)^2$ and $(p_i+p_j+p_k)^2$
respectively. The $z_i$ are energy fractions defined wrt the original
parton's energy so that we have $\sum_i z_i=1$. Also, in what follows
below we shall need only the splitting functions in four space-time
dimensions and hence have set the dimensional regularisation parameter
$\epsilon$ to zero above and in all subsequent applications.

The phase-space in the triple-collinear limit can be written as
\begin{equation}
d \Phi_3 = \frac{\left(p_t R \right)^4}{\pi}\left(z_1 z_2 z_3 \right) dz_2 \, dz_3 \, d\theta_{12}^2 \, d\theta_{23}^2 \,
d\theta_{13}^2  \Delta^{-1/2} \Theta\left(\Delta\right),
\end{equation}
with the Gram determinant $\Delta$ given by
\cite{GehrmannDeRidder:1997gf,Bertolini:2015pka}
\begin{equation}
\Delta = 4 \theta_{13}^2
\theta_{23}^2-(\theta_{12}^2-\theta_{13}^2-\theta_{23}^2)^2.
\end{equation}

We then carry out an integral over the triple-collinear phase-space
which includes the action of the taggers encoded as a sequence of
kinematical cuts. We compute the jet mass distribution as an
integral of the schematic form

\begin{equation}
\label{eq:fulldistbn}
\left(\frac{\rho}{\sigma}  \frac{d\sigma}{d\rho}\right)^{\mathrm{LO,triple-collinear}}
  = \left(\frac{\alpha_s}{2\pi}\right)^2
   \int  d\Phi_3 \, 
  \frac{\langle \hat{P} \rangle}{s_{123}^2} \,
  \Theta^{\mathrm{jet}} \,
  \Theta^{\mathrm{tagger}}(\zetacut,\rho_{\mathrm{min}})\, \rho \, \delta \left(\rho- \frac{s_{123}}{R^2 p_t^2}
\right), \,
\end{equation}
where $\langle \hat{P} \rangle $ denotes the spin-averaged
triple-collinear splitting function, including the proper symmetry
factor for identical particles, for the splitting of an initial quark
(or gluon if considering a gluon-initiated jet), the
$\Theta^{\mathrm{jet}}$ condition denotes the constraint for all three
partons to be in the same anti-$k_t$ jet of a given radius $R$
\begin{equation}\label{eq:antikt_3partons}
\Theta^{\mathrm{jet}} = \sum_{i>j \neq k} 
  \Theta \Big(d^{\text{(anti-$k_t$)}}_{ij}<\mathrm{min}\big(d^{\text{(anti-$k_t$)}}_{ik},d^{\text{(anti-$k_t$)}}_{jk}\big)\Big)\,
     \Theta(\theta_{ij}<R) \Theta(\theta_{(i+j)k}<R),
\end{equation}
and the condition $\Theta^{\mathrm{tagger}}$ represents the action of
the substructure taggers. In particular, $\Theta^{\mathrm{tagger}}$
contains constraints from the $\zetacut$ and $\rhomin$ conditions
which will regulate the soft and collinear divergences of the
$1 \to 3$ splitting functions.
Accordingly we can carry out the computation of the jet mass
distribution entirely in 4 dimensions and only real-emission terms
contribute at the leading order $\alpha_s^2$.

For any of the taggers we have introduced, we have, at order $\alpha_s^2$,
\begin{multline}\label{eq:theta_tagger_alphasqr-base}
\Theta^{\text{tagger}} \left(\zetacut,\rhomin \right)
   = \sum_{i>j \neq k} 
  \Theta \left(d^{\text{(tagger)}}_{ij}<\mathrm{min}(d^{\text{(tagger)}}_{ik},d^{\text{(tagger)}}_{jk})\right)\,
     \Theta \left (\mathrm{min}(z_k,1-z_k) > \zetacut \right) \times \\
\times \Theta \left(\mathrm{min}(z_i,z_j) >  \zetacut \right) \,
 \Theta \left(\mathrm{min}(\rho_{ij},\rho_{jk},\rho_{ki}) > \rhomin \right),
\end{multline}
where the only difference between the CMS (recall that there is no
difference at order $\alpha_s^2$ between the default CMS,
$\mathrm{CMS^{3p,mass}}$ and \TopSplitter) and the \Ym-splitter
taggers is in the distance measure they use:
\begin{align}
d^{\text{(CMS)}}_{ij}         & = \theta_{ij}^2,\\
d^{\text{(\Ym-splitter)}}_{ij} & = \text{min}(z_i,z_j)\theta_{ij}^2.
\end{align}
However, at this order of the perturbation theory,
Eq.~\eqref{eq:theta_tagger_alphasqr-base} is equivalent to the simpler
form
\begin{equation}\label{eq:theta_tagger_alphasqr}
\Theta^{\text{tagger}} \left(\zetacut,\rhomin \right)
   =  \Theta \left(\mathrm{min}(z_1,z_2,z_3) >  \zetacut \right) \,
 \Theta \left(\mathrm{min}(\rho_{12},\rho_{13},\rho_{23}) > \rhomin \right),
\end{equation}
which is the same for the CMS and \Ym-splitter taggers.

It is worth noting that the triple-collinear splitting functions and
phase-space are \emph{not} presently included in current parton shower
models implemented in any of the main general purpose Monte Carlo
event generator codes.\footnote{For recent attempts at partially
  including these effects in parton showers we refer the reader to
  \cite{Hoche:2017iem}.}
Parton showers instead include the strongly-ordered limit of the
triple-collinear functions where the triple-collinear functions
factorise into a product of two leading-order splitting kernels. It is
simple to make this link explicit by expanding the triple-collinear
functions about the strongly ordered limit. For instance for the
$C_F^2$ term reported in Eq.~ \eqref{eq:CF2} we can explicitly take
the limit $\theta_{23}^2 \ll \theta_{13}^2$ and perform an expansion
in the smallest angle $\theta_{23}^2$. Writing
$\theta_{12}^2 = \theta_{13}^2+\theta_{23}^2-2\theta_{13} \theta_{23}
\cos\phi$
and introducing the splitting variables $z$ and $z_p$ such that
$z_1 = 1-z, z_2=z(1-z_p), z_3 = z z_p$, one obtains upon series
expansion in $\theta_{23}^2$ :
\begin{equation}
\frac{{\langle \hat{P} } ^{(\mathrm{ab})}_{g_1 g_2 q_3}\rangle}{s_{123}^2}d\Phi_3
  = C_F^2 \frac{d\theta_{13}^2}{\theta_{13}^2}
          \frac{d\theta_{23}^2}{\theta_{23}^2}\Theta(\theta_{23}<\theta_{13})
          \,dz \,dz_p \, \frac{d\phi}{2 \pi}
    \bigg( \frac{1+z^2}{1-z} \times \frac{1+z_p^2}{1-z_p}
    + \mathcal{O} \left(\theta_{23} \right) \bigg).
\end{equation}
The above form exhibits the factorisation of the leading order
splitting kernels which is expected in the strongly ordered
limit.\footnote{In general i.e.\ beyond the Abelian $C_F^2$ term a
  fully factorised structure is obtained \emph{after} an azimuthal
  integration in the strongly ordered limit.}  Additionally taking the
soft limit i.e.\ $z, z_p \to 1$ brings us back to the approximations
used to derive \eqref{eq:cmsll}.

\section{Resummed calculation to all orders}\label{sec:analytic-resum}

Eq.~(\ref{eq:cmsll2}), making use of the $1\to 3$ splitting function
to obtain $f_q(\rho,\rhomin,\zetacut)$, is sufficient to obtain the
small $\rho$ (and $\rhomin$) behaviour at order $\alpha_s^2$. 
However, the large logarithms of $\rho$ or $\rhomin$ need to be
resummed to all orders in $\alpha_s$. 
For this, in addition to the two hard collinear emissions described by
the $1\to 3$ splitting, we need to add an arbitrary number of real or
virtual soft and collinear emissions and consider the constraints on
them.
As is standard in resummation, this is expected to yield a Sudakov
form factor that multiplies the leading-order result.
In this section, we derive the explicit form of that Sudakov for the
\TopSplitter, $\mathrm{CMS^{3p,\mathrm{mass}}}$ and \Ym-splitter top
taggers. The IRC unsafety of the default CMS tagger prevents a similar
analysis being directly carried out for that case.

Before digging into the details of each tagger, let us clarify the
accuracy of our resummation. First of all, previous work has
shown~\cite{Dasgupta:2013ihk,Dasgupta:2015lxh,Dasgupta:2015yua,Dasgupta:2016ktv}
that a leading-logarithmic (LL) calculation is usually sufficient to
grasp the main features of substructure tools.
In that context, our resummation should definitely include large
logarithms of the jet mass ($\rho$ or $\rhomin$) to LL accuracy. These
logarithms are the most relevant for describing boosted jets and we
shall see that including them holds the key to understanding the basic
details of the top-taggers.

However, since both $\rhomin/\rho$ and $\zetacut$ are somewhat smaller
than unity for practical applications, it might also be of interest to
include logarithms in $\rhomin/\rho$ and $\zetacut$ in the
resummation. Indeed, without including these terms one may worry about
their impact on our analytical picture for top tagging.
With this in mind our resummation accuracy goal will ideally be
double-logarithmic, but where the scale of the logarithms can be
either $\rho$ (or $\rhomin$), $\rhomin/\rho$ or $\zetacut$. We will
also confirm that in the final result the logarithms of $\rho$ or
$\rhomin$ dominate over logarithms of $\rhomin/\rho$ and $\zetacut$ in
the region relevant for phenomenology, as one might have naively
expected purely on grounds of their numerical size.

While our final resummation accuracy is double-logarithmic, or more
precisely leading-logarithmic after inclusion of running coupling
effects, we shall also retain some sources of single-logarithmic
corrections, notably via the inclusion of hard-collinear contributions
which arise from considering the full splitting functions rather than
just their soft-enhanced terms.  This is again standard in the
existing resummed calculations for jet substructure (see
e.g.\ Ref.~\cite{Dasgupta:2013ihk}) and is sometimes referred to as
modified leading logarithmic accuracy~\cite{Larkoski:2014wba}.

We would like to stress that strictly from the point of view of our
logarithmic accuracy, we do not need the full structure of the
triple-collinear splitting functions that we have used above to
compute the leading-order pre-factor, that will eventually multiply
the Sudakov exponent. Instead one could just treat the pre-factor in
the soft limit with strong angular-ordering, which is sufficient to
retain all double-logarithmic terms in the pre-factor. Using the
triple-collinear splitting functions means that we have instead chosen
to be more careful in our treatment of the pre-factor by retaining
terms that are formally subleading from the viewpoint of our
logarithmic resummation accuracy. In effect we thus perform a form of
matching so that at order $\alpha_s^2$ our result coincides with using
the full triple-collinear splitting function, while beyond order
$\alpha_s^2$ our result should contain all potentially large
double-logarithmic terms, counting logs of
$\rho,\rho_{\mathrm{min}}, \rho/\rho_{\mathrm{min}}$, and $\zetacut$
on the same footing. In practice we are able to achieve this goal for
the \TopSplitter and \Ym-splitter taggers including also \Ym-splitter
with general SoftDrop gre-grooming. Instead for the case of the
$\mathrm{CMS^{3p,\mathrm{mass}}}$ tagger it does not prove to be
simple to include logarithms of $\zetacut$ and
$\rho/\rho_{\mathrm{min}}$ on the same footing as double logarithms in
$\rho$ or $\rho_{\mathrm{min}}$. Accordingly for
$\mathrm{CMS^{3p,\mathrm{mass}}}$ we do not attempt to retain all
possible double logarithms, focusing mainly on the numerically
dominant logarithms in $\rho$ or $\rho_{\mathrm{min}}$. This level of
accuracy is still sufficient to provide us insight into the behaviour
of the tagger in the region relevant for phenomenology.

The structure and details of the resummed result depend on the tagger
being considered. We first discuss the \Ym-splitter case due to its
simpler structure.

\subsection{\Ym-splitter}\label{sec:resum-ysplitter}

At double-logarithmic accuracy, we can consider emissions to be soft
and strongly ordered, here in gen-$k_t$ distance, or, equivalently, in
mass. If one wishes to simultaneously retain the information that is
present in the triple-collinear limit however, we have to lift the
requirement of strong ordering and the soft approximation, for the two
emissions that are declustered by the taggers, while still retaining
these approximations for all remaining emissions.  In the first
instance however it is instructive to impose the soft and
strong-ordered requirement on {\emph{all}} emissions including the
declustered ones, which gives us the leading-logarithmic accuracy we
seek. Subsequently we shall match our result to the triple-collinear
limit.

\subsubsection{Calculation in pure soft and strongly-ordered
  limit}\label{sec:resum-Ysplit-strongordering}

We first consider two soft emissions $k_1$ and $k_2$ both emitted by a
hard parton leg with colour factor $C_R$, where $C_R =C_F$ for a quark
initiated jet and $C_R=C_A$ for a gluon initiated jet. We denote by
$x_i$ and $\theta_i$ the momentum fraction and angle of emission $k_i$
defined wrt the hard emitting parton. At leading logarithmic level we
can assume strong ordering in the gen-$k_t$ distance or equivalently
in masses.  Hence we assume
$\rho\approx \rho_1\equiv x_1\theta_1^2\gg \rho_2 \equiv
x_2\theta_2^2$, which implies that emission $k_1$ is the first to be
declustered and $k_2$ is the second, while $k_1$ also sets the jet
mass $\rho$.

Next, we consider multiple soft emissions ordered in gen-$k_t$
distance. Emissions $k_1$ and $k_2$ are, by construction, the ones
obtained by the declustering procedure and subject to the
$\zetacut$ requirement. Due to the gen-$k_t$ ordering they are also
the emissions that dominate the pairwise masses entering the
$\rho_{\mathrm{min}}$ condition.\footnote{Technically, the gen-$k_t$
  distance between prong $i$ and prong $j$ differs from the pairwise
  mass in two ways: first by an overall factor proportional to
  $z_i+z_j$, and second by factors of the form $1-z$ which are
  irrelevant in the resummation limit $z\ll 1$. Overall, this means
  that the relative ordering of the emissions in mass is the same as
  their relative gen-$k_t$ ordering.}
Thus all the tagger constraints are fully determined by the
declustered partons $k_1$ and $k_2$ which produce the leading-order
pre-factor, cf. Eq.~(\ref{eq:cmsll}).

\begin{figure}
\centering
\includegraphics[width=0.32\textwidth]{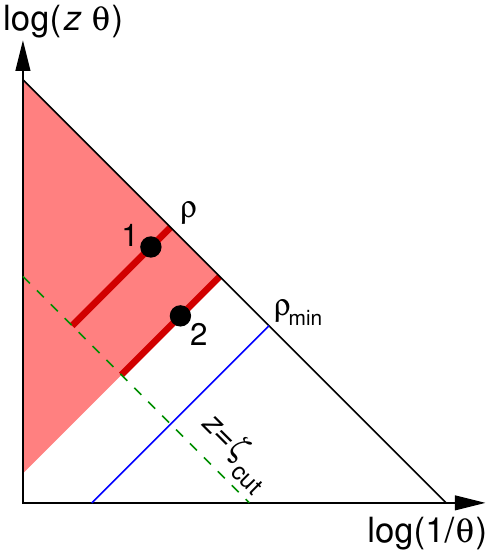}%
\caption{Lund plane corresponding to the \Ym-splitter tagger for top
  tagging. The emission that has a larger gen-$k_t$ distance or mass
  is labelled 1 and that with the next largest mass is labelled
  2. Emissions 1 and 2 both pass the $\zetacut$ condition shown using
  the dashed line. The mass $\rho_2$ lies between $\rho$ and
  $\rho_{\mathrm{min}}$ as shown. The red shaded region represents the
  region over which emissions are vetoed and leads to the appearance
  of the primary emission contribution to the Sudakov form factor. A
  similar configuration where the second-largest mass emission, 2, is
  emitted as a secondary emission from emission 1, is not shown in
  this plot.}
\label{fig:lund-ysplit}
\end{figure}

One then has a veto on any additional emission with gen-$k_t$
distance (or, equivalently, mass) larger than $\rho_2=x_2\theta_2^2$.
For (primary) emissions from the leading parton $p$, this corresponds
to the shaded (red) region in the plot of
Fig.~\ref{fig:lund-ysplit}. In this region, virtual emissions are
still allowed, yielding a Sudakov form factor
$\exp(-R_{\text{\Ym-splitter}}^{\text{(primary)}})$, with
$R_{\text{\Ym-splitter}}^{\text{(primary)}}$ corresponding to the
shaded area:
\begin{equation}\label{eq:RYsplitter-primary}
R_{\text{\Ym-splitter}}^{\text{(primary)}}(\rho_2) = 
  \int \frac{d\theta^2}{\theta^2} \frac{dz}{z}\,
  \frac{\alpha_s(z\theta p_tR)C_R}{\pi}\,
  \Theta(z\theta^2>\rho_2),
\end{equation}
where we also took into account running coupling effects.

Furthermore, since $\rho_2\ll \rho_1$, we should also veto (secondary)
emissions from $k_1$ in between these two scales.  Consider such an
emission from $k_1$ to be soft and to carry a momentum fraction
$z \ll 1$ of the momentum of its parent $k_1$ implying it has momentum
fraction $x_1 z$ wrt the jet $p_t$.  It is emitted at angles smaller
than $\theta_1$ due to angular ordering. Such emissions should also be
vetoed if they give a gen-$k_t$ distance larger than
$\rho_2$.

This is not shown in Fig.~\ref{fig:lund-ysplit} and gives an
additional contribution (with $x_1=\rho_1/\theta_1^2$)
\begin{equation}\label{eq:RYsplitter-secondary}
R_{\text{\Ym-splitter}}^{\text{(secondary)}}(\rho_2; \rho_1, \theta_1) = 
  \int \frac{d\theta^2}{\theta^2} \frac{dz}{z}\,
  \frac{\alpha_s(zx_1\theta p_tR)C_A}{\pi}\,
  \Theta(zx_1\theta^2>\rho_2)\,\Theta(\theta<\theta_1).
\end{equation}
Note that in the expression for secondary emissions, the ordering in
gen-$k_t$ distance (imposed by the declustering procedure of
\Ym-splitter), differs from the mass by an $x_1$ factor.\footnote{Note
  that it is still true that inside the $k_1$ prong, i.e.\ amongst all
  the secondary emissions off $k_1$, the emission that has the largest
  gen-$k_t$ distance also dominates the mass of the prong.}

Ultimately, the Sudakov form factor is
\begin{align}\label{eq:SYsplitter}
{\cal S}_{\text{\Ym-splitter}}(\rho_2; \rho_1, \theta_1) 
 & = \exp\Big[-R_{\text{\Ym-splitter}}(\rho_2; \rho_1, \theta_1)\Big],\\
R_{\text{\Ym-splitter}}(\rho_2; \rho_1, \theta_1)
 & = R_{\text{\Ym-splitter}}^{\text{(primary)}}(\rho_2)
   + R_{\text{\Ym-splitter}}^{\text{(secondary)}}(\rho_2; \rho_1, \theta_1).
\end{align}

In the strongly-ordered (and soft) limit, the resummed result,
including the pre-factor is therefore
\begin{multline}\label{eq:resum-Ysplitter-strongly-ordered}
\left(\frac{\rho}{\sigma} \frac{d\sigma}{d\rho}\right)^{\text{resum}}
  = \int_0^1 \frac{d\theta_1^2}{\theta_1^2}
  \frac{d\theta_2^2}{\theta_2^2}
  \frac{dx_1}{x_1}\frac{dx_2}{x_2} \, 
  \frac{\alpha_s(x_1\theta_1p_tR)C_R}{\pi} \frac{\alpha_s(x_2\theta_2p_tR)C_R}{\pi} \\
  \Theta(x_1>\zetacut)\,
  \Theta(x_2>\zetacut)\,
  \Theta(\rhomin<x_2\theta_2^2<x_1\theta_1^2) \,
  \rho \, \delta(\rho-x_1\theta_1^2)\,{\cal S}_{\text{\Ym-splitter}}.
\end{multline}
The above result coincides with the LO result in Eq.~(\ref{eq:cmsll})
at order $\alpha_s^2$, i.e.\ when switching off the running of the
strong coupling and the Sudakov form factor and replacing $C_R$ by
$C_F$. As part of our accuracy goal which aims at correctly retaining
\emph{all} double logarithms, our pre-factor should also contain the
$\mathcal{O} \left(\ln^3 \zeta_{\mathrm{cut}} \right)$ terms neglected
in Eq.~(\ref{eq:cmsll}). Moreover we should account for all possible
branchings that contribute to the pre-factor including the case where
emission $k_2$ is emitted as a secondary emission off $k_1$, which for
a quark initiated jet yields a $C_F C_A$ contribution to the
pre-factor. However since, in the next subsection, we eventually use
the triple-collinear splitting functions and kinematics to compute our
pre-factor, it is guaranteed that all such terms (along with
subleading terms relevant beyond the soft and strongly ordered
approximation) are correctly retained.

Below we give results in the fixed-coupling case to highlight the
different logarithms that are present in the above expressions:
\begin{align}
R_{\text{\Ym-splitter}}^{\text{(primary)}}(\rho_2) 
  & \overset{\text{f.c.} }{=} \frac{\alpha_sC_R}{2\pi}\ln^2\rho_2, \\  
\label{eq:sec} R_{\text{\Ym-splitter}}^{\text{(secondary)}}(\rho_2; \rho_1, \theta_1)
  & \overset{\text{f.c.}}{=} \frac{\alpha_sC_A}{2\pi} \ln^2(\rho_2/\rho).
\end{align}

After integration over $\rho_2$, the numerically dominant logarithms
will be of the form of a series in $\alpha_s\ln^2\rho$ where we treat
$\rho$ and $\rho_{\mathrm{min}}$ on the same footing, which multiplies
the leading order pre-factor and originates purely from the veto on
primary emissions. Given the range of the $\rho_2$ integration between
$\rho_{\mathrm{min}}$ and $\rho$, secondary emissions can only
contribute terms which are at most as singular as
$\alpha_s\ln^2(\rho/\rhomin)$. We may therefore anticipate that
secondary emissions turn out to be relatively significant only when
$\rho_{\mathrm{min}} \ll \rho$, which is an observation we shall
return to later.

While the fixed-coupling results we have reported here (and the
corresponding results derived for other taggers later in this section)
are computed purely in the soft limit, our final results include not
just the running coupling effects but also the impact of hard
collinear corrections via inclusion of the ``$B_1$" resummation
coefficients associated to the splitting kernels, which are beyond our
formal double logarithmic accuracy.
The full expressions, including running-coupling effects and
hard-collinear splittings are given in Appendix~\ref{sec:radiators}.

\subsubsection{Matching to the triple-collinear limit}\label{sec:resum-Ysplit-matching}

As we argued at the end of section~\ref{sec:analytic-leading}
(cf. Eq.~(\ref{eq:cmsll2})), a more accurate calculation of the
pre-factor multiplying the Sudakov form factor involves lifting the
assumption of strong ordering between the two leading emissions $k_1$
and $k_2$ (and that of their softness). One should then use the
$1\to 3$ splitting function for calculating the pre-factor. 
This is best described using the kinematics of
section~\ref{sec:analytic-triple}, i.e.\ a system of 3 partons $p_1$,
$p_2$ and $p_3$, carrying jet momentum fractions $z_i$, with
$\sum_i z_i=1$, and with pairwise angles $\theta_{ij}$.

Matching this to the resummed results obtained in
section~\ref{sec:resum-Ysplit-strongordering} comes with two
conditions. First, we need to make sure that the emissions on which
the Sudakov depends (i.e.\ $p$, $k_1$ and $k_2$ in the previous
section), which are by construction the emissions on which the two
$\zetacut$ constraints are imposed, are also the relevant emissions
that are constrained by the $\rhomin$ condition imposed on the
pre-factor, cf. the $\Theta^{\text{tagger}}$ factor in
Eq~(\ref{eq:fulldistbn}). Once this condition is satisfied, we will need
to map the emissions $p$, $k_1$ and $k_2$ onto the triple-collinear
parton system.

For the first condition, as already mentioned, the fact that we are
using a gen-$k_t$ declustering guarantees that the emissions picked by
the declustering procedure are also the emissions that dominate the
pairwise prong masses.

We are therefore left with mapping the emissions $p$, $k_1$ and $k_2$
onto the triple-collinear system. Within our double-logarithmic
accuracy, this is equivalent to redefine the arguments of the Sudakov
($\theta_1$, $\rho_1$ and $\rho_2$) in terms of the kinematics of
partons $p_1$, $p_2$ and $p_3$ to account for the lifting of the
strong-ordering and softness assumptions.
The Sudakov form factor can then still be formally written as
Eqs.~(\ref{eq:RYsplitter-primary}) and
(\ref{eq:RYsplitter-secondary}). However there is a freedom in
defining $\theta_1$, $\rho_1$ and $\rho_2$ since our only constraint
is to recover the proper soft and strongly-ordered limit.

Ultimately, the all-order version of Eq.~(\ref{eq:fulldistbn}) becomes
\begin{equation}\label{eq:resum-Ysplitter}
\frac{\rho}{\sigma} \frac{d\sigma}{d\rho}
  = \int  d\Phi_3 \, 
  \frac{\langle \hat{P} \rangle}{s_{123}^2} \,
  \frac{\alpha_s(k_{t1})}{2\pi} \frac{\alpha_s(k_{t2})}{2\pi}  \Theta^{\mathrm{jet}} \,
  \Theta^{\mathrm{tagger}}\, \delta
  \left(\rho- \frac{s_{123}}{R^2 p_t^2}\right)\,{\cal
    S}_{\text{\Ym-splitter}}(\rho_2; \rho_1, \theta_1),
\end{equation}
where we still have to specify $\theta_1$, $\rho_i$ and $k_{ti}$.
For definiteness, let us consider the situation where the partons
$p_2$ and $p_3$ are clustered first, followed by a clustering of the
$(p_2,p_3)$ pair with $p_1$ for which we adopted the following
prescription:
\begin{align}
\theta_1 & = \theta_{1(2+3)}, &
\theta_2 & = \theta_{23},\\
\rho_1 & = \text{min}(z_1,1-z_1)\theta_1^2, & 
\rho_2 & = \text{min}(z_2,z_3)\theta_2^2,\label{eq:def-rhos-ysplitter}\\
k_{t1} & = \text{min}(z_1,1-z_1)\theta_1 p_tR, & 
k_{t2} & = \text{min}(z_2,z_3)\theta_2 p_tR,\\
x_1 & \equiv \rho_1/\theta_1^2 =  \text{min}(z_1,1-z_1).
\end{align}
which can be easily verified to agree with the resummed expressions
above, in the soft and strongly-ordered limit. It is perhaps worth
re-emphasising that using the form of a double-logarithmic Sudakov
form factor multiplying the triple-collinear splitting functions
produces uncontrolled terms beyond our leading-logarithmic
accuracy. The main purpose of introducing the triple-collinear
splitting is to calculate the order $\alpha_s^2$ pre-factor as
precisely as possible while beyond
$\mathcal{O} \left(\alpha_s^2\right)$ only double logarithmic terms
are controlled.

\subsubsection{\Ym-splitter with grooming}\label{sec:Ysplit-grooming}

In Ref.~\cite{Dasgupta:2015yua} it was shown that Y-splitter was a
high performance tool for tagging two-pronged decays only when
supplemented with grooming e.g.\ via mMDT \cite {Dasgupta:2013ihk} or
trimming \cite{Krohn:2009th}. It was also shown that the order in
which Y-splitter and grooming were used on the jet was crucial to the
performance. Grooming jets \emph{after} using Y-splitter resulted in a
subleading impact on the crucial large Sudakov suppression of the QCD
background seen with Y-splitter, hence maintaining this desirable
feature. On the other hand grooming significantly increased the signal
efficiency over that seen with plain Y-splitter. The improved signal
efficiency and largely unmodified background suppression resulted in
striking gains to the signal significance ($S/\sqrt{B}$). On the other
hand using grooming tools such as trimming and mMDT \emph{before}
Y-splitter was seen in comparison to not give a good performance ,
since the background Sudakov suppression factor changed from the
Y-splitter Sudakov to the less effective trimming and mMDT Sudakov
factors respectively. An exception to this, noted in
Ref.~\cite{Dasgupta:2016ktv}, was SoftDrop pre-grooming with positive
$\beta$ (typically, $\beta=2$), where grooming had a smaller impact on
the Y-splitter Sudakov, while the signal efficiency and sensitivity to
non-perturbative effects were still considerably improved. In this
respect of achieving a high performance while minimising
non-perturbative effects, the SoftDrop ($\beta=2$) pre-grooming option
followed by \Ym-splitter emerged as one of the most effective and
reliable methods in the analysis of Ref.~\cite{Dasgupta:2016ktv}.

In the present case, i.e.\ for top tagging, it shall turn out to be
interesting to explore both $\beta=0$ (i.e.\ mMDT) and $\beta>0$
pre-grooming options. Indeed based on our previous work
\cite{Dasgupta:2016ktv} we may anticipate that pre-grooming with mMDT
produces a Sudakov that resembles the mMDT Sudakov suppression
factor. As we shall show in the next section, this is also the
essential behaviour shown by the CMS tagger (when one considers our
IRC safe extensions thereof).

\begin{figure}
\centering
\begin{subfigure}{0.32\textwidth}
\includegraphics[width=\textwidth]{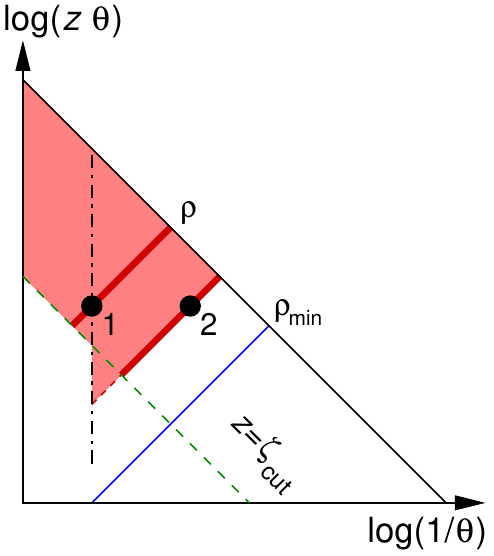}\caption{}\label{fig:lund-ysplit-mMDT1}%
\end{subfigure}\hfill
\begin{subfigure}{0.32\textwidth}
\includegraphics[width=\textwidth]{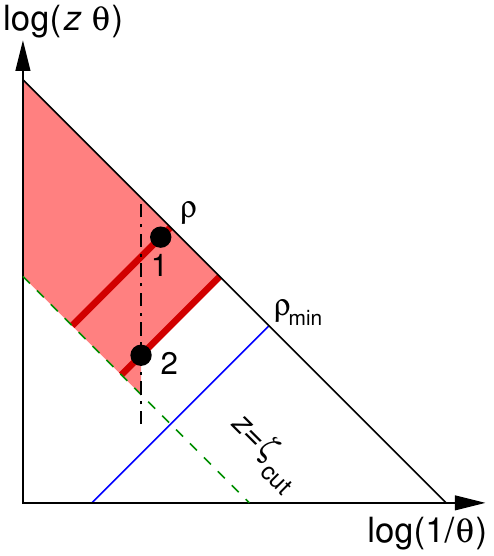}\caption{}\label{fig:lund-ysplit-mMDT2}%
\end{subfigure}\hfill
\begin{subfigure}{0.32\textwidth}
\includegraphics[width=\textwidth]{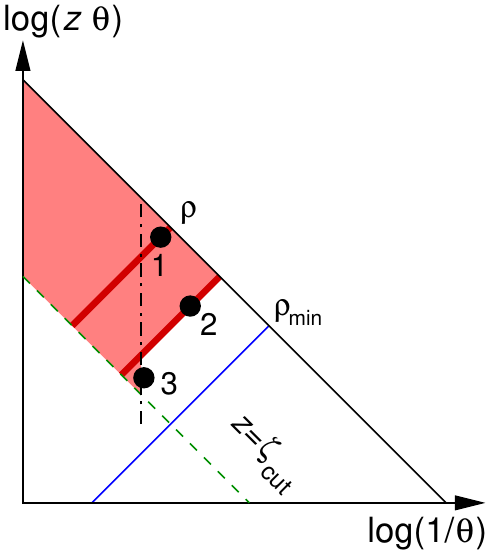}\caption{}\label{fig:lund-ysplit-mMDT3}%
\end{subfigure}
\caption{Lund plane corresponding to \Ym-splitter with
  grooming. Emissions 1 and 2 have respectively the largest and
  second-largest gen-$k_t$ distance (or mass). The 3 plots (a), (b)
  and (c) correspond to three different possibilities where the
  emissions 1, 2 and 3 respectively are the largest-angle emissions
  that pass the mMDT condition.}
\label{fig:lund-ysplit-mMDT}
\end{figure}

We first consider SoftDrop pre-grooming for $\beta=0$ i.e.\ with
mMDT. After applying mMDT to the jet we apply \Ym-splitter as adapted
by us for top-tagging (see section.~\ref{sec:Ymsplittertagger-def} for
details). For simplicity we use the $\zetacut$ condition with the same
value for both grooming and \Ym-splitter. One needs to consider three
separate cases represented in Fig.~\ref{fig:lund-ysplit-mMDT}:
\begin{enumerate}
\item The largest angle emission that passes mMDT is also the largest
  gen-$k_t$ (or equivalently largest mass) emission from those that
  remain after grooming (Fig.~\ref{fig:lund-ysplit-mMDT1}). This
  emission is also the first to be declustered by \Ym-splitter and
  sets the final jet mass $\rho$.
\item The largest angle emission that passes mMDT is the second
  largest gen-$k_t$ (mass) emission and is hence the second emission
  to be declustered by \Ym-splitter (Fig.~\ref{fig:lund-ysplit-mMDT2}).
\item The largest angle emission that passes mMDT is lower in mass
  than either of the two emissions declustered by \Ym-splitter
  (Fig.~\ref{fig:lund-ysplit-mMDT3}). This situation first occurs at
  the level of three real emissions i.e.\ at order $\alpha_s^3$.
\end{enumerate}

To obtain the result corresponding to the first case, consider the
emissions $k_1$ with largest mass $z_1 \theta_1^2 =\rho_1=\rho$ and
$k_2$ with second largest mass $z_2 \theta_2^2 = \rho_2$ as before,
with $\theta_1 > \theta_2$ and with $z_1,z_2 >\zetacut$.  The first
emission is by assumption the emission that passes mMDT which means
that all emissions larger in angle that fail the $\zetacut$ condition
have been groomed away. Here the mMDT stops and one applies
\Ym-splitter to the jet, declustering $k_1$ and $k_2$ and imposing the
$\rho_{\mathrm{min}}$ conditions as before. The condition that there
are no emissions with mass larger than $\rho_2$ after mMDT (except the
real emission $k_1$) would produce the standard mMDT Sudakov in
$\rho_2$. However in the present case a key difference with mMDT is
the fact that mMDT stops at emission $k_1$.  This implies that
emissions $k_i$ at angles smaller than $\theta_1$ that have
$z_i < \zetacut$ are no longer removed by mMDT. If such emissions set
a mass larger than $\rho_2$ they must be vetoed as well which gives
an extra contribution to that arising from the mMDT Sudakov down to
$\rho_2$ (as explicitly visible in
Fig.~\ref{fig:lund-ysplit-mMDT1}). The result for the overall Sudakov
exponent for the primary emission contribution may be written as
\begin{equation}\label{eq:radiator-ysplit-mmdt-case1}
  R^{(1),\text{primary}}(\rho_2;\rho_1,\theta_1)
  = R_{\text{mMDT}}(\rho_2) +R_{\text{mMDT}}^{\text{angle}}(\theta_1,\rho_2),
\end{equation}
where $R^{(1)}$ represents the Sudakov that applies in the situation
that the emission that passes the mMDT condition is also the largest
mass emission, $R_{\text{mMDT}}(\rho_2)$ is the standard mMDT Sudakov
down to mass $\rho_2$ and $R_{\text{mMDT}}^{\text{angle}}(\theta_1,\rho_2)$ is the
contribution from vetoing emissions that are at smaller angles than
$\theta_1$, have $z<\zetacut$ and set a mass larger than $\rho_2$. A
straightforward calculation in the fixed-coupling approximation gives
\begin{equation}\label{eq:radiator-ysplit-mmdt-angle}
R_{\text{mMDT}}^{\text{angle}}(\theta_1,\rho_2) = \frac{C_R \alpha_s}{2 \pi} \left [ \ln^2 \frac{\zetacut \theta_1^2}{\rho_2} \Theta \left(\zetacut \theta_1^2 > \rho_2 \right) \right],
\end{equation}
while the mMDT Sudakov is the usual known result~\cite{Dasgupta:2013ihk}
\begin{equation}\label{eq:radiator-ysplit-mmdt-mmdt}
R_{\text{mMDT}}(\rho_2)  = \frac{C_R \alpha_s}{2\pi} \left[\Theta \left(\zetacut> \rho_2\right) \left(2 \ln \frac{1}{\zetacut} \ln \frac{1}{\rho_2}-\ln^2 \frac{1}{\zetacut}\right) +\Theta \left(\rho_2>\zetacut \right) \ln^2 \frac{1}{\rho_2} \right].
\end{equation}

For the second case where the emission that passes mMDT is $k_2$ i.e
the second largest in mass the mMDT evolution down to $\rho_2$ is
unmodified and hence we obtain
\begin{equation}
\label{eq:radiator-ysplit-mmdt-case2}
R^{(2),\text{primary}}(\rho_2;\rho_1,\theta_1) = R_{\text{mMDT}}(\rho_2).
\end{equation}

Finally in the third case where an emission $k_3$ triggers mMDT before
either $k_1$ or $k_2$, for such an emission to be allowed its mass
should be smaller than the mass set by $k_2$. This contribution
cancels against corresponding virtual corrections as for the case of
the standard mMDT calculation (corresponding to the small triangular
areas with mass smaller than $\rho_2$ in
Figs.~\ref{fig:lund-ysplit-mMDT2}
and~\ref{fig:lund-ysplit-mMDT3}). Such configurations can thus be
ignored.

In addition to the primary emission contributions considered above,
there is also a secondary emission contribution to the
Sudakov. Secondary emission contributions are not modified by mMDT
pre-grooming and hence in either of the two cases considered above,
the secondary emission result coincides with that already obtained for
\Ym-splitter in Eq.~\eqref{eq:sec}.

Therefore ultimately, the background distribution can still be written
in the form of Eq.~(\ref{eq:resum-Ysplitter}) now with a primary
Sudakov given by Eqs.~(\ref{eq:radiator-ysplit-mmdt-case1})
and~(\ref{eq:radiator-ysplit-mmdt-case2}).
Note that, on top of showing different Sudakov suppressions, the
different kinematic cases from Fig.~\ref{fig:lund-ysplit-mMDT} will
also be weighted differently when inserted in
Eq.~(\ref{eq:resum-Ysplitter}).

The main feature of the mMDT pre-grooming and \Ym-splitter is that the
result closely resembles the known mMDT result i.e.\ one inherits the
Sudakov structure of the pre-grooming method. At small jet masses
$\rho_2$, the Sudakov has an $\alpha_s \ln \zetacut \ln \rho_2$
behaviour which gives a smaller suppression than the
$\alpha_s \ln^2 \rho_2$ behaviour of \Ym-splitter.  Differences from
the pure mMDT result arise due to the extra
$R_{\text{mMDT}}^{\text{angle}}$ term and due to secondary
emissions. In both cases the argument of the double logarithm produced
has a ratio involving either $\theta_1^2/\rho_2$ or $\rho/\rho_2$
which again can be expected to be modest contributions, except
possibly at small values of $\rho_{\mathrm{min}}$.

For the case of SoftDrop pre-grooming using a general $\beta >0$ the
same general arguments hold as for the $\beta=0$ mMDT results derived
above. The only change one needs to make is that the condition for an
emission to pass the SoftDrop constraint now becomes
$z>\zetacut \theta^{\beta}$. Hence we get
\begin{equation}
R_{\text{SD}}^{\text{angle}}(\theta_1,\rho_2,\beta) = \frac{C_R\alpha_s}{2 \pi} \frac{2}{2+\beta} \ln^2 \frac{\zetacut \theta_1^{2+\beta}}{\rho_2} \Theta\left(\zetacut
  \theta_1^{2+\beta} > \rho_2\right)
\end{equation}
while $R_{\mathrm{mMDT}}(\rho_2)$ is replaced by the SoftDrop Sudakov
down to $\rho_2$:
\begin{equation}\label{eq:RSD}
R_{\text{SD}}(\rho_2,\beta)  = \frac{C_R \alpha_s}{2\pi} \left [ \Theta(\zetacut>\rho_2)  \left ( \ln^2 \frac{1}{\rho_2}-\frac{2}{2+\beta} \ln^2\frac{\zetacut}{\rho_2} \right)+ \Theta \left(\rho_2 > \zetacut \right)\ln^2 \frac{1}{\rho_2}  \right ].
\end{equation}
The primary-emission Sudakov therefore becomes
\begin{equation}\label{eq:radiator-ysplit-mmdt}
  R_{\text{SD+\Ym-splitter}}^{\text{(1),primary}}(\rho_2;\rho_1,\theta_1)
  = R_{\text{SD}}(\rho_2) +R_{\text{SD}}^{\text{angle}}(\theta_1,\rho_2),
\end{equation}
while we also have as for the mMDT pre-grooming case the contribution
$R_{\text{SD+\Ym-splitter}}^{\text{(2),primary}}(\rho_2) =
R_{\text{SD}} (\rho_2)$.

As for the case of \Ym-splitter, our final results also
include running-coupling effects and hard-collinear splittings, and
are given in Appendix~\ref{sec:radiators}.

\subsection{\TopSplitter and CMS$^\mathrm{3p,\mathrm{mass}}$}
\label{sec:resum-cms}
As discussed before and explicitly demonstrated in
Appendix~\ref{app:event2}, the CMS tagger is unsuitable for precise
theoretical computations involving top jets, due to its IRC unsafety
at high $p_t$ which also rules out all-orders perturbative
calculations such as those based on resummation, for the CMS
tagger. Instead we shall consider our extensions of the tagger
i.e.\ the CMS$^{\mathrm{3p,\mathrm{mass}}}$ variant and the method we
call \TopSplitter which also originates from the CMS tagger. In fact
the \TopSplitter method turns out to be the most amenable to a
resummed calculation to the accuracy we were able to achieve for
\Ym-splitter and \Ym-splitter with grooming, namely the resummation of
logarithms of $\rho/\rho_{\mathrm{min}}$ and $\zetacut$ on the same
footing as logarithms of $\rho$ or $\rho_{\mathrm{min}}$. Hence we
shall consider this method first.

For the \TopSplitter tagger we shall need to consider Cambridge-Aachen
(C-A) declustering of the jet rather than the gen-$k_t$ declustering
relevant for \Ym-splitter.  Therefore now we need to consider emissions
that are soft and strongly ordered in angle. In particular if we
assume that emission $k_1$ is declustered first then there is a veto
on any emission at an angle larger than $\theta_1$ with $z > \zetacut$
(emissions with $z<\zetacut$ are groomed away by the primary
declustering procedure). On declustering the emission $k_1$ we are
left with $k_1$ and a massive prong $p$.  The tagger then proceeds to
decluster prong $p$.  The declustering produces a second emission
$k_2$, also with $z > \zetacut$, which, by definition of the
\TopSplitter method, dominates the mass of the prong and hence we
impose a veto on all emissions that set a larger mass than
$\rho_2 = z_2 \theta_2^2.$\footnote{Note that the fact that the
  declustered emission is the one that dominates the prong mass owes
  precisely to our construction of \TopSplitter which picks the
  emission with largest $p_{ti} \theta_i^2$ as a product of the
  declustering.} The veto on emissions in the prong is only active for
emissions with $z> \zetacut$. To see this, note that emissions in the
prong at angles larger than that of emission $k_2$ and with
$z<\zetacut$ are groomed away by the secondary declustering step of
the tagger, while emissions with angles smaller than that of $k_2$ and
with $z<\zetacut$ cannot dominate the mass in any case.
Furthermore, with the above \TopSplitter procedure of selecting $k_2$
so that it dominates the prong mass, only emissions $k_1$ and $k_2$
enter into the construction of the minimum pairwise mass and
contribute to the pre-factor that multiplies the Sudakov
exponent.

\begin{figure}
\centering
\begin{subfigure}{0.48\textwidth}\centering
\includegraphics[width=0.66\textwidth]{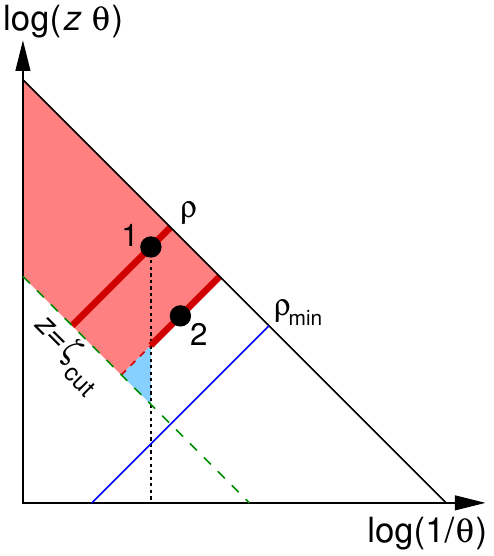}\caption{}\label{fig:lund-cms-case1}%
\end{subfigure}
\hfill
\begin{subfigure}{0.48\textwidth}\centering
\includegraphics[width=0.66\textwidth]{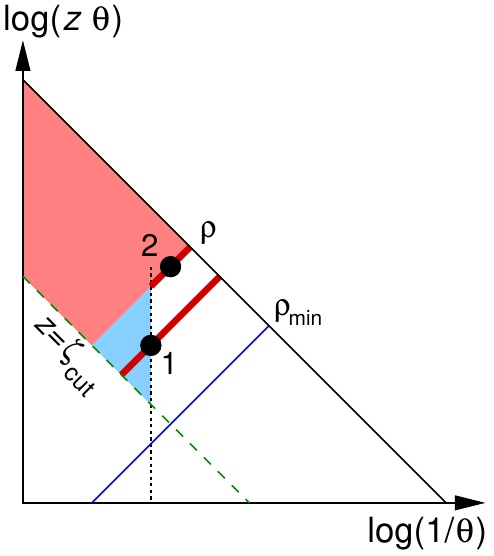}\caption{}\label{fig:lund-cms-case2}%
\end{subfigure}
\caption{Lund plane corresponding to the \TopSplitter tagger.
  Emissions are labelled such that emission 1 is the first one
  selected by the tagger, i.e.\  $\theta_1\gg \theta_2$, and emission
  2 dominates the prong mass. The two plots correspond to (a)
  $\rho_1\gg\rho_2$ and (b) $\rho_1\ll\rho_2$.
  The red shaded region shows the veto region for emissions with
  $z>\zetacut$ and $\rho >\rho_2$ while the blue shaded triangle
  region corresponds to the extra contribution that arises from
  requiring that there are also no emissions with $\theta > \theta_1$
  and $z> \zetacut$.}\label{fig:lund-cms} 
\end{figure}

The situation is depicted in the Lund plane in
Fig.~\ref{fig:lund-cms}. Emissions $k_1$ and $k_2$ are shown
corresponding to $\theta_1 \gg \theta_2$ with either $\rho_1 \gg
\rho_2$ (Fig.~\ref{fig:lund-cms-case1}) or $\rho_1 \ll
\rho_2$ (Fig.~\ref{fig:lund-cms-case2}).
The red shaded region shows a veto on all emissions with mass larger
than $\rho_2$ and $z > \zetacut$ as argued above. A further blue
shaded region shows additional emissions that are vetoed since they
have $\theta>\theta_1$ and $z > \zetacut$. 
Analogous to the case of \Ym-splitter we can write a resummed result
of the form given in Eq.~\eqref{eq:resum-Ysplitter} but with a
different Sudakov form factor $\mathcal{S}_{\mathrm{\TopSplitter}}$
which can be written as
\begin{equation} 
\mathcal{S}_{\mathrm{\TopSplitter}} = \exp \left[-R_{\text{\TopSplitter}} \left(\rho_2;\rho_1,\theta_1 \right) \right].
\end{equation}

The Sudakov exponent $R_{\mathrm{\TopSplitter}}$ receives
contributions from both vetoes on primary and secondary emissions
i.e.\ $R_{\text{\TopSplitter}} =
R_{\text{\TopSplitter}}^{\text{(primary)}}+R_{\text{\TopSplitter}}^{\text{(secondary)}}$.
The veto on primary emissions was discussed above and its explicit
form is:
\begin{equation}\label{eq:RCMS-primary}
R_{\text{\TopSplitter}}^{\text{(primary)}}(\rho_2;\rho_1,\theta_1) = 
  \int \frac{d\theta^2}{\theta^2} \frac{dz}{z}\,
  \frac{\alpha_s(z\theta p_tR)C_R}{\pi}\,
  \Theta(z>\zetacut)\,
  \Theta(z\theta^2>\rho_2 \text{ or }\theta>\theta_1).
\end{equation}

We should also consider the case of secondary emissions from $k_1$
which would prevent emission $k_2$ from being declustered if they have
mass larger than $\rho_2$ and energy fraction wrt the jet $p_t$
greater than $\zetacut$, hence we must also veto such emissions. In
this case we obtain
\begin{equation}\label{eq:RCMS-secondary}
R_{\text{\TopSplitter}}^{\text{(secondary)}}(\rho_2; \rho_1, \theta_1) = 
  \int^{\theta_1^2} \frac{d\theta^2}{\theta^2} \frac{dz}{z}\,
  \frac{\alpha_s(zx_1\theta p_tR)C_A}{\pi}\,
  \Theta(x_1 z> \zetacut)\,
  \Theta(zx_1^2\theta^2>\rho_2).
\end{equation}
In the above equation $z$ and $\theta$ are the energy fraction and
angle of the secondary emission with respect to the emitting parent
$k_1$, which itself has energy fraction $x_1$ and angle $\theta_1$ with
respect to the hard jet $p_t$ and direction respectively. Also, as
  for the \Ym-splitter case discussed in
  section~\ref{sec:resum-ysplitter}, one could also have a situation
  where emission $k_2$ is emitted as a secondary emission from
  $k_1$. Again, this situation is automatically accounted for by
  matching to the triple-collinear splitting function.
The main features of the results, for the
above defined contributions, are again best illustrated by using a
fixed-coupling approximation and we refer to
Appendix~\ref{sec:radiators} for our full expressions including
running-coupling results and hard-collinear splittings. It
is instructive to further separate the contributions to
$R_{\text{\TopSplitter}}^{\text{(primary)}}$ and write it as the sum
of the contributions due to the red shaded region ($z\theta^2>\rho_2$)
and the extra blue shaded region ($z\theta^2<\rho_2$ and
$\theta>\theta_1$) in Fig.~\ref{fig:lund-cms},
$R_{\text{\TopSplitter}}^{\text{(primary)}} =
R_{\text{\TopSplitter}}^{\text{(red)}}+
R_{\text{\TopSplitter}}^{\text{(blue)}}$.  For the red shaded region
we obtain just the usual result for the mMDT already mentioned in
Eq.~\eqref{eq:radiator-ysplit-mmdt-mmdt},
\begin{equation}
R_{\text{\TopSplitter}}^{\text{(red)}}(\rho_2)  = R_{\text{mMDT}}(\rho_2),
\end{equation}
 while the blue triangle contributes as below:
\begin{equation}
 R_{\text{\TopSplitter}}^{\text{(blue)}}(\rho_2;\theta_1)  =
\frac{\alpha_sC_R}{2\pi}  \left [ \ln^2 \frac{\rho_2}{\zetacut
    \theta_1^2}\, \Theta(\rho_2>\zetacut\theta_1^2) - \ln^2 \frac{\rho_2}{\zetacut}\, \Theta(\rho_2>\zetacut) \right ].
\end{equation}

The corresponding expression for
$R_{\text{\TopSplitter}}^{\text{(secondary)}}$ in a fixed-coupling
approximation is also simple to obtain:

\begin{equation}
R_{\text{\TopSplitter}}^{\text{(secondary)}}(\rho_2; \rho_1, \theta_1) = \frac{C_A \alpha_s}{2 \pi} \left[ \ln^2 \frac{x_1 \rho}{\rho_2} \Theta \left(\rho_2 < x_1 \rho \right)-\ln^2 \frac{\zetacut \rho}{\rho_2} \Theta \left(\rho_2 < \zetacut \rho \right) \right].
\end{equation}

Some comments about the results obtained here are in order. Although
we have identified various different contributions, the most relevant
contribution to the tagger behaviour comes from the
$R_{\text{\TopSplitter}}^{\text{(red)}}$ term which is essentially the
same Sudakov as was originally obtained for the modified mass-drop
tagger (mMDT/SD($\beta=0$))~\cite{Dasgupta:2013ihk}. This
is because as we mentioned before the largest logarithms are those in
jet mass, and in the limit of small jet mass $\rho_2 \ll \zetacut$, we
see a single-logarithmic Sudakov suppression due to
$R_{\text{\TopSplitter}}^{\text{(red)}}$. The remaining terms i.e those
due to secondary emissions and
$R_{\text{\TopSplitter}}^{\text{(blue)}}$ produce, in the limit of small
jet mass, only leading logarithms of $\zetacut$ and $\rho_1/\rho_2 $ or
equivalently $\rho/\rho_{\mathrm{min}}$. We retain these terms here
for the reasons mentioned before, namely to assess their impact on the
tagger behaviour.

Lastly we are left with mapping the variables entering the Sudakov
onto the triple-collinear set of emissions $p_1,p_2,p_3$ as for the
\Ym-splitter case. We again exploit our freedom to choose the precise
definitions, with the only constraint being to recover the
leading-logarithmic results after taking the soft and strongly-ordered
limits. We then define (again for the case where $p_2$ and $p_3$ are
clustered first followed by $p_1$ with the $(p_2,p_3)$ pair):
\begin{align}
\rho_1 & = z_1(1-z_1)\theta_1^2, & \rho_2 & = z_2z_3\theta_2^2,\label{eq:def-rhos-topsplitter} \\
k_{t1} &= z_1(1-z_1) \theta_1 p_tR, & k_{t2} &= z_2 z_3 \theta_2 p_tR, \\
\theta_1 & = \theta_{1(2+3)}, &
\theta_2 & = \theta_{23}.
\end{align}

Having discussed the case of \TopSplitter, we now turn to the
$\mathrm{CMS^{3p,\mathrm{mass}}}$ tagger. The main difference with the
\TopSplitter case is simply the fact that when declustering a prong,
one takes the largest angle emission within the prong, i.e the one
declustered first by the tagger, as a product of the
declustering. This emission is not guaranteed to dominate the mass of
the prong however, and hence there is a possible mismatch between the
emissions that are declustered by the tagger and those that enter the
minimum pairwise mass condition, in particular for the pairwise mass
that involves one of the branches from the secondary declustering and
the branch left intact from the primary declustering. Configurations
for which there is such a mismatch produce additional corrections with
leading-log terms involving logs of $\rho_{\mathrm{min}}/\rho$ and
$\zetacut$. In this case, the resummation of such terms is possible
but substantially more complicated than for the \TopSplitter
case. Since the behaviour of the $\mathrm{CMS^{3p,\mathrm{mass}}}$
tagger now depends on up to three emissions (two emissions dominating
the $\zetacut$ condition and one additional emission dominating the
$\rho_\text{min}$ condition), the matching with the triple-collinear
splitting is no longer systematically achievable.
Given the small impact of the additional terms, both numerically and
for our understanding of the tagger behaviour, we will neglect
them. Hence, in our analytical treatment, the result for the
$\mathrm{CMS^{3p,\mathrm{mass}}}$ variant of the CMS tagger coincides
with that we presented for \TopSplitter. We shall later verify that
the performance of $\mathrm{CMS^{3p,\mathrm{mass}}}$ and \TopSplitter,
as given by parton shower models, is in fact consistent with our
observations.

\section{Results}\label{sec:results}

We have previously noted that given the fact that $\rho$ and
$\rho_{\mathrm{min}}$ are not widely disparate in practice, the
approximation of strong angular-ordering may not be sufficient to
satisfactorily capture the impact of top taggers on QCD jets. It is
thus clearly interesting to attempt to compare the results obtained
using the triple-collinear splitting functions to those produced by
the strong angular-ordering approximation, especially since it is the
latter picture that is effectively included in most parton shower
descriptions of QCD jets. At the same time given that we have devoted
most of this article to discussing resummation effects in detail, it
is also worthwhile to consider the numerical importance of resummation
and especially to understand the relative contributions of various
different contributions to the Sudakov exponents. We shall devote the
current section to these studies.

\subsection{Numerical impact of triple-collinear and resummation
  effects}\label{sec:resum-summary}

\begin{figure}
  \centering
  \begin{subfigure}{0.49\textwidth}
  \includegraphics[width=\textwidth]{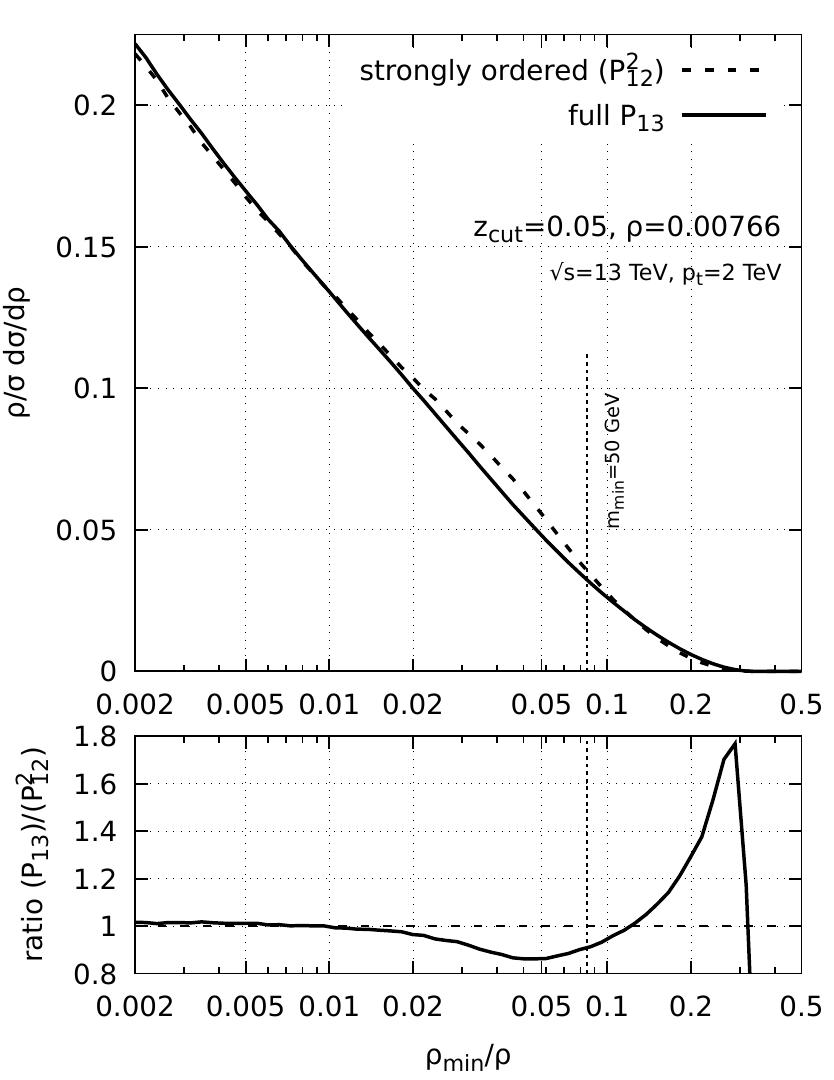}
  \caption{no resummation}\label{fig:effect13-nosudakov}
  \end{subfigure}
  \begin{subfigure}{0.49\textwidth}
  \includegraphics[width=\textwidth]{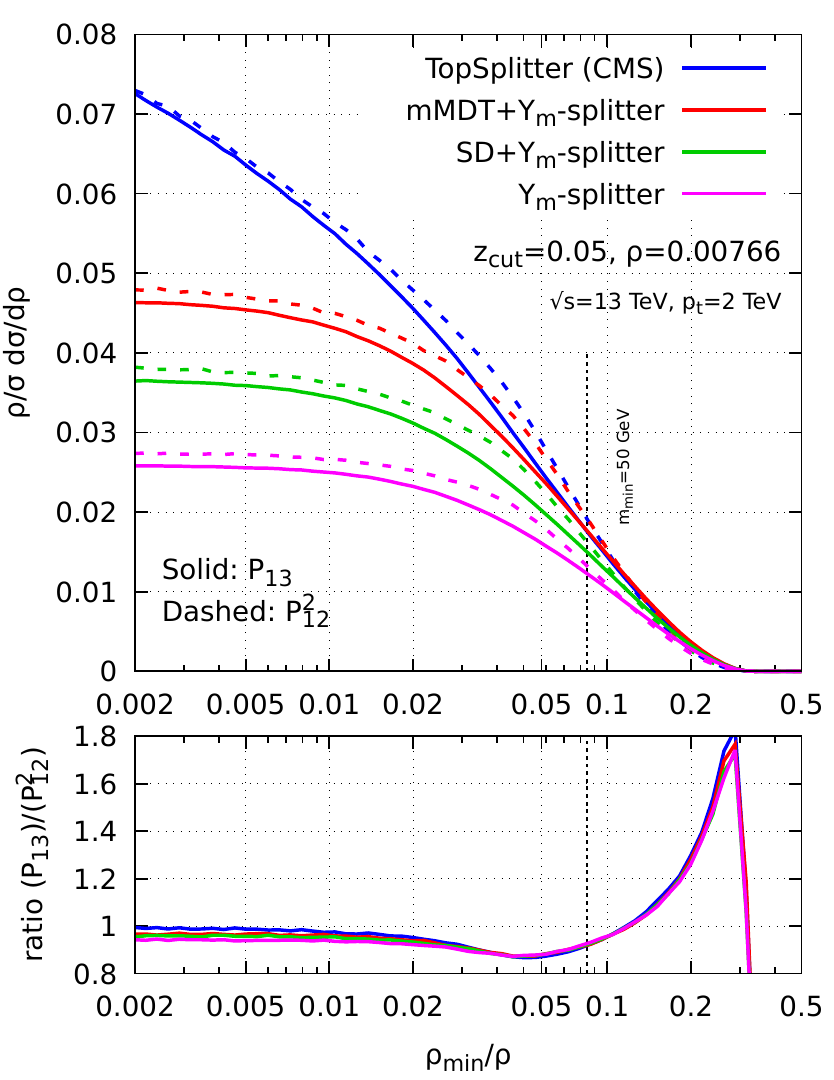}
  \caption{with resummation}\label{fig:effect13-sudakov}
  \end{subfigure}
  \caption{Comparison of the results obtained for quark jets in the
    strongly-ordered limit (dashed curves) with the results using the
    full triple-collinear splitting function (solid curves), (a)
    without including resummation effects, (b) including resummation
    effects. In both cases, the top panel shows the distributions
    $\rho/\sigma\,d\sigma/d\rho$ and the lower panel shows the ratio
    between the results including the full triple-collinear splitting
    and the strongly-ordered limit.}\label{fig:effect13}
\end{figure}

We first discuss the effect of including the full triple-collinear
splitting function instead of working in the strongly-ordered limit.
This is shown in Fig.~\ref{fig:effect13} both with and without inclusion of 
resummation effects.
As expected, in the limit $\rho_\text{min}\ll\rho$, both results
agree, although the ratio does not exactly converge to 1 in the case
where resummation effects are included simply because the Sudakov form
factor weights differently different regions of phase-space. 
For situations closer to what is used for phenomenology,
i.e.\ $m_{\text{min}}\approx 50$~GeV (highlighted by the vertical
dotted line on the plots), the inclusion of the full
triple-collinear splitting function only introduces a correction of
about 10\% once all effects are considered.
This means that unless one uses larger values of $\rho_\text{min}$,
closer to the endpoint of the distribution, the effect of the
triple-collinear splitting functions is modest and should not
substantially modify pure parton shower descriptions of top tagging
mistag rates.

\begin{figure}
  \includegraphics[width=0.5\textwidth,page=1]{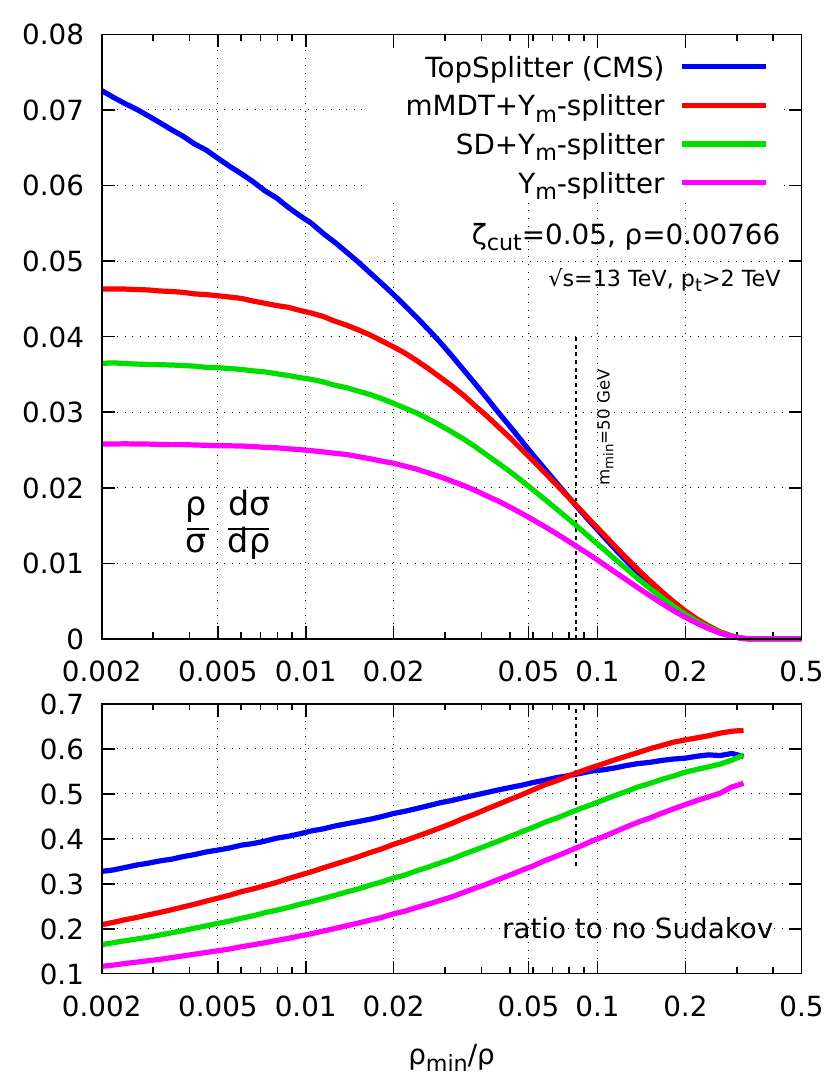}\hfill
  \includegraphics[width=0.5\textwidth,page=2]{figs/sudakov-choices.pdf}
  \caption{Study of the resummation effects for the various taggers as
    a function of $\rhomin$. The top-left plot shows the distribution
    $\rho/\sigma d\sigma/d\rho$, and the bottom-left plot shows the
    overall Sudakov effect (the ratio of $\rho/\sigma d\sigma/d\rho$
    with and without the Sudakov form factor). On the right, different
    levels of approximation to the Sudakov were made: (top) simply
    using the jet-mass Sudakov (plain, SD or mMDT depending on the
    tagger) down to the scale $\rho$, (middle) considering instead the
    full Sudakov from primary emissions, and (bottom) adding secondary
    emissions. Each plot shows the ratio to the previous
    approximation.}\label{fig:sudakov-choices}
\end{figure}

Next, we move on to the discussion of resummation effects.
We have plotted in Fig.~\ref{fig:sudakov-choices} our results for
$\rho/\sigma d\sigma/d\rho$, obtained from (\ref{eq:resum-Ysplitter})
adapted to each tagger, varying $\rhomin$.
The plot shows the overall effect of the resummation on the left and
the effects split in a series of contributions on the right. Focusing
on the left plot first, we see that the resummation has a
sizeable impact, suppressing the QCD background by a factor $\sim
2-3$, in the phenomenological region. As expected, the effects increase
when further reducing $\rhomin$.

The series of plots on the right of Fig.~\ref{fig:sudakov-choices} aim
at gauging the relative importance of various types of contribution to the
Sudakov.
Here we studied 4 different levels of approximations for the Sudakov
form factor. First, we generated results without a Sudakov form factor
(i.e.\ with just the leading-order $\alpha_s^2$ calculation with the
$1\to 3$ splitting function), then those with just a simplified
Sudakov exponent involving resummation of only logarithms of $\rho$
via the radiator $R(\rho)$. $R(\rho)$ is taken as the plain jet-mass
radiator for the case of \Ym-splitter, the appropriate groomed
jet-mass radiator for \Ym-splitter with grooming and the mMDT radiator
for \TopSplitter.
Next, we studied results involving only primary emissions and finally
the full result including all double logarithms on the same footing
and including secondary emissions.
The three plots show the ratio of the results obtained with one
approximation relative to what was obtained with the previous (more
crude) approximation. 

The top plot shows the effect of the jet-mass like Sudakov
$\exp(-R(\rho))$, compared to not including any Sudakov.
This is expected to capture the dominant logarithms, i.e.\ the most
enhanced by logarithms of $\rho$, in the phenomenological region. We
see indeed that they come with a large suppression factor.
Furthermore, we see that the suppression is larger for \Ym-splitter
than for SD+\Ym-splitter, itself more suppressed than
mMDT+\Ym-splitter and \TopSplitter, following the size of the region
vetoed by the Sudakov factor.

In the middle plot, we now include the full primary Sudakov (recall
that the plot shows the relative impact of the full primary Sudakov
compared to just including ``$\exp(-R(\rho))$'').
Although this is expected to have a smaller effect than the
resummation of the dominant logarithms of $\rho$, typically trading a
logarithm of $\rho$ for a logarithm of $\rhomin/\rho$, we see that the
effect remains sizeable, in fact, almost as large as the
$\exp(-R(\rho))$ Sudakov.
Again, decreasing $\rhomin$ the effect of the full primary Sudakov
becomes more pronounced, dominating the trend seen at small $\rhomin$
in the overall Sudakov effect (bottom-left plot).
As before, the Sudakov suppression is reduced when the level of
pre-grooming is increased.
Note that, although this is not explicitly shown in the plot, we have
also tested the relative importance of the ``blue'' primary Sudakov
compared to the (dominant) ``red'' contribution in the case of
\TopSplitter and found that it had a very small effect of order of a
few percent at most.

Finally, the bottom-right plot studies the effect of adding the
secondary emission suppression, shown as the ratio of the results with
the full Sudakov compared to only including primary emissions. This is
expected to involve only additional logarithms of $\rhomin/\rho$ and
$\zetacut$ and it indeed turns out to have a small impact in the
region relevant for phenomenology, again increasing when moving to
smaller values of $\rhomin$.

Before moving to a comparison to parton-shower Monte-Carlo
simulations, we note that in the strongly-ordered limit and using a
fixed-order approximation for the Sudakov, it is possible to simplify
(at least some of) the integrations over emissions 1 and 2 in
(\ref{eq:resum-Ysplitter}).
For the case of \Ym-splitter, all the integrations can be performed
analytically. The full analytic result explicitly highlights the expected
logarithmic dependences and nicely reproduces the various trends
observed in Fig.~\ref{fig:sudakov-choices}.
For other taggers, we could only obtain interesting expressions in the
limit $\rhomin/\rho\ll\zetacut\ll\rho$ or
$\zetacut\ll\rhomin/\rho\ll\rho$ which again involved the expected
dominant logarithms and corroborated the behaviours seen numerically
in Fig.~\ref{fig:sudakov-choices}.

\subsection{Comparison to parton showers}
\label{sec:analytic-v-shower}

\begin{figure}
  \includegraphics[width=0.33\textwidth,page=1]{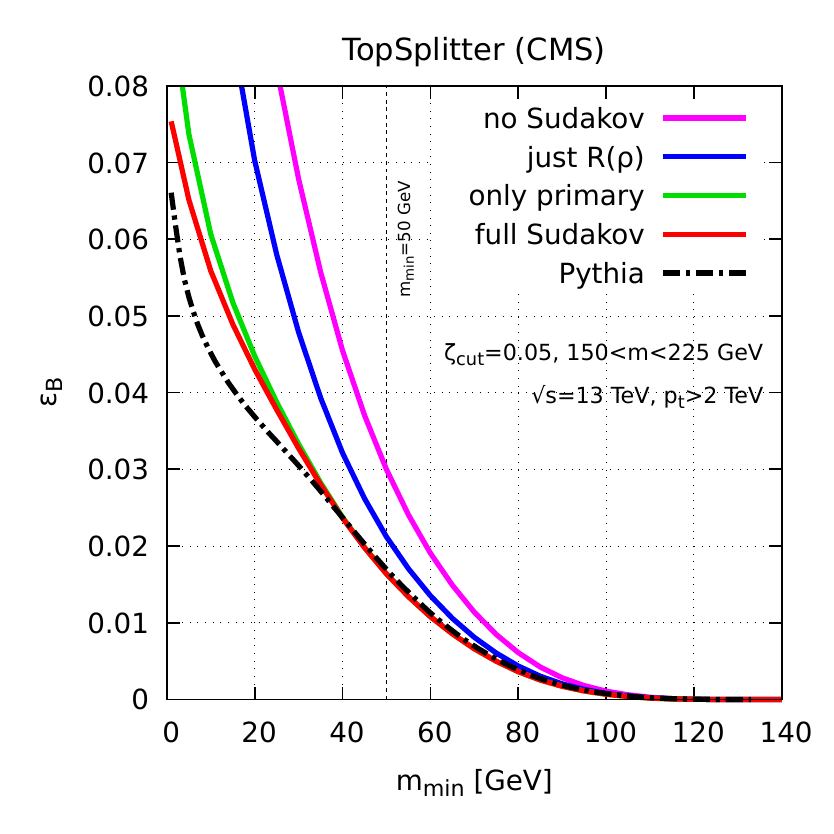}\hfill
  \includegraphics[width=0.33\textwidth,page=2]{figs/sudakov-v-mc.pdf}\hfill
  \includegraphics[width=0.33\textwidth,page=4]{figs/sudakov-v-mc.pdf}
  \caption{Comparison between analytic results and Pythia simulations
    for the QCD background efficiency. Results are plotted as a
    function of the $m_\text{min}$ cut. For the analytic results, we have
    included the same levels of approximation to the Sudakov as in
    Fig.~\ref{fig:sudakov-choices}.}\label{fig:sudakov-v-mc}
\end{figure}

Having obtained analytic results for the different taggers, in this
section we shall compare the analytics to Monte Carlo (MC)
simulations. Here we will be interested in parton level MC results,
since we are comparing to a purely perturbative calculation, i.e.\ we
shall use Pythia 8.230~\cite{Sjostrand:2014zea} parton-level events to
compare to our all-order resummed analytic predictions. The impact of
non-perturbative effects (hadronisation and MPI) will be considered
when we discuss tagger performances in the next section.

We consider jet production in dijet processes at the LHC with
$\sqrt{s}=13 \, \mathrm{TeV}$. We first focus on subprocesses
involving only quark jets in the final state (by selecting $qq\to qq$
hard matrix elements) and will discuss gluon jets (obtained through
the $gg\to gg$ matrix element) later. We define jets using the
anti-$k_t$ algorithm~\cite{Cacciari:2008gp}, as implemented in
FastJet~\cite{Cacciari:2005hq,Cacciari:2011ma}, and use a jet radius
$R=1$ and a transverse momentum selection cut such that
$p_t > 2 \, \mathrm{TeV}$. For all the taggers we use $\zetacut =0.05$. To
study the background efficiency (mistag rate) of the taggers we work
in a mass window around the signal mass
$150 < m < 225 \, \mathrm{GeV}$.

First we compare in Fig.~\ref{fig:sudakov-v-mc} our analytic
predictions to parton shower results for the background efficiency,
obtained by integrating over the signal mass window, as a function of
$m_{\mathrm{min}}$.
We consider the case of \TopSplitter for which the LL resummation
structure resembles that for the $\mathrm{CMS^{3p,\mathrm{mass}}}$
variant of the CMS tagger but where we control all double logarithms
and not just those in $\rho$, as would be the case for
$\mathrm{CMS^{3p,\mathrm{mass}}}$. We also consider both \Ym-splitter
alone as well as its combination with pre-grooming via mMDT and
SoftDrop ($\beta=2$).
For each tagger we show the analytic results using the same four
levels of approximation to the Sudakov as for
Fig.~\ref{fig:sudakov-choices}. We also show the result from Pythia
for comparison.

Let us first consider the \TopSplitter results. We note that the best
agreement across all $m_\mathrm{min}$ values is provided by the use of
the full Sudakov. In the phenomenologically relevant region with
$m_{\mathrm{min}} \sim 50$ GeV one obtains perfect agreement with
Pythia by using the full Sudakov while using $R(\rho)$ alone in the
Sudakov exponent gives a noticeable difference with Pythia which
increases at small $m_{\mathrm{min}}$.  At smaller $m_{\mathrm{min}}$
beyond the phenomenologically relevant region, one sees
that Pythia starts to depart somewhat from the analytic
resummation. The feature in the Pythia results at small
$m_{\mathrm{min}}$ is not evident in the analytic calculations but
occurs in a region where the pure parton shower predictions are potentially
subject to significant non-perturbative corrections.  To see this one
notes that jet masses $\sim 40$ GeV can be produced by emissions with
energy $\sim 1$ GeV in conjunction with a hard parton with energy
$\sim 2 $ TeV. Hence the difference between Pythia's parton shower
(without hadronisation) and analytics at such low masses is largely of
academic interest. As already observed in
Fig.~\ref{fig:sudakov-choices}, secondary emissions have only a modest
role over most of the $m_{\mathrm{min}}$ range though at smaller
$m_{\mathrm{min}}$ there is evidence that they have the effect of
shifting the resummed result closer towards those from Pythia.

Next we discuss the plain \Ym-splitter case. We again observe a good
general agreement of the full resummed result with Pythia across a
broad range of $m_{\mathrm{min}}$ with some difference visible at
smaller $m_{\mathrm{min}}$ values somewhat beyond the
phenomenologically relevant values. Secondary emissions play a more
visible role here at smaller $m_{\mathrm{min}}$ than for \TopSplitter
and noticeably move the result closer to that from Pythia.  Similar
comments apply to the groomed variants of \Ym-splitter with again a
good general agreement for the full resummed result with Pythia and a
demonstrable improvement from including resummation effects beyond
those in the naive $R(\rho)$ function.

\begin{figure}
  \includegraphics[width=0.5\textwidth]{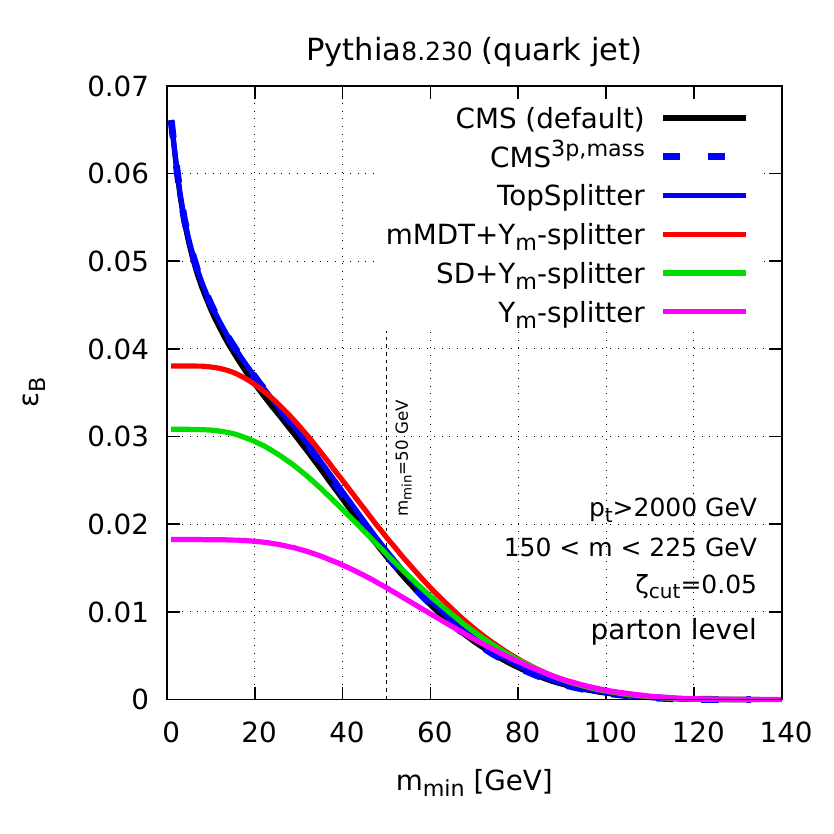}
  \hfill
  \includegraphics[width=0.5\textwidth]{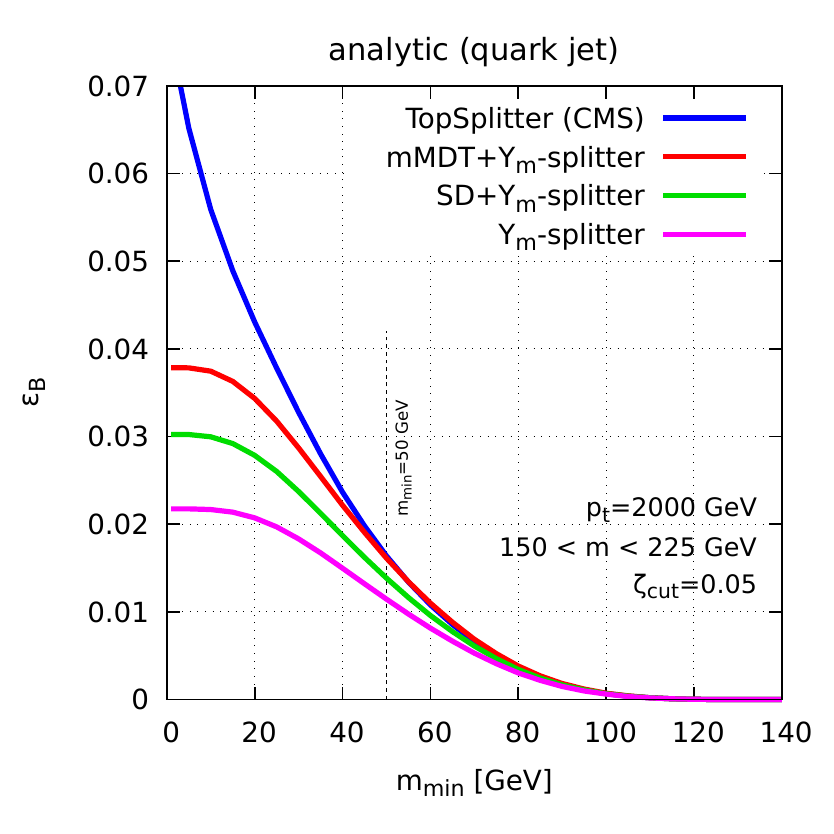}
  \caption{Background efficiency as a function of the $m_{\text{min}}$
    cut for various taggers.
    The left plot shows results obtained from Monte Carlo simulations
    with the Pythia8 generator and the right plot shows the results of
    our analytic calculations discussed in the main text.
  }\label{fig:mmin-distrib}
\end{figure}

For ease of comparison between the different taggers, we also show in
Figure \ref{fig:mmin-distrib} results for the background efficiency or
mistag rate $\epsilon_B$ of the different taggers as a function of
$m_{\mathrm{min}}$ on the same plot, with MC results shown on the left
and analytic results on the right.
Taggers with a lower $\epsilon_B$ suppress the background more, which
is desirable, although the final performance depends also on the
impact of the taggers on signal jets, which we discuss in the next
section.
As far as the main purpose of this section is concerned --- comparing
expectations from analytics with results from MC parton showers ---
one can say that the general features of the MC results are well
reproduced by the analytics. In particular one notes the ordering in
the performance of taggers that is predicted by the analytics also
emerges in the parton shower results. We would naturally expect, as
has also been observed before~\cite{Dasgupta:2016ktv} for the case of
two-pronged signal jet substructure, that \Ym-splitter suppresses the
background most effectively due to the large double-logarithmic
Sudakov form factor obtained there. This expectation is clearly borne
out by both the analytical and MC results. We would also expect that
\Ym-splitter with pre-grooming using SoftDrop ($\beta=2$) would give
the SoftDrop Sudakov which reduces the background less than
\Ym-splitter but still more than other methods with a smaller Sudakov
suppression. This also emerges in the MC studies albeit at somewhat
smaller $m_{\mathrm{min}}$ than predicted by the analytics. Next one
would expect the mMDT pre-groomed \Ym-splitter which has an
essentially mMDT style Sudakov suppression (a smaller suppression than
that expected from SoftDrop ($\beta=2$)) however still retaining the
\Ym-splitter result at the level of secondary emissions. Again,
especially at slightly smaller $m_{\mathrm{min}}$ relative to the
analytics, the MC results follow this expected trend. Lastly we have
the case of \TopSplitter where relative to the \Ym-splitter based
methods one would expect the smallest Sudakov suppression since both
primary and secondary emissions are impacted by the $\zetacut$
condition.  Once again MC results confirm this expectation.

Perhaps most crucially, at the phenomenological working point of
$50~\mathrm{Gev}$ there is no significant difference visible in the
analytics between the results for \TopSplitter and those for
mMDT+\Ym-splitter and this is also what emerges in the parton shower
results. A small difference is visible between the above two methods
and SoftDrop ($\beta=2$) in the analytics while the spread in MC
results is not visible yet. Finally there is a clear difference
between the above three methods and \Ym-splitter visible in both
analytics and MC.

A further comment is due on the MC results for the original CMS
tagger, labelled as CMS (default) in Figure~\ref{fig:mmin-distrib}
compared to those for the $\mathrm{CMS^{3p,\mathrm{mass}}}$ variant
and \TopSplitter. The MC predictions for these methods are in
remarkably good agreement with one another, being virtually coincident
over the entire $m_{\mathrm{min}}$ range. This suggests that both
$\mathrm{CMS^{3p,\mathrm{mass}}}$ and \TopSplitter are good IRC safe
alternatives to the CMS tagger, with \TopSplitter having the advantage
of being more amenable to an accurate resummation of all
double-logarithmic enhanced terms.

\begin{figure}
  \includegraphics[width=0.5\textwidth]{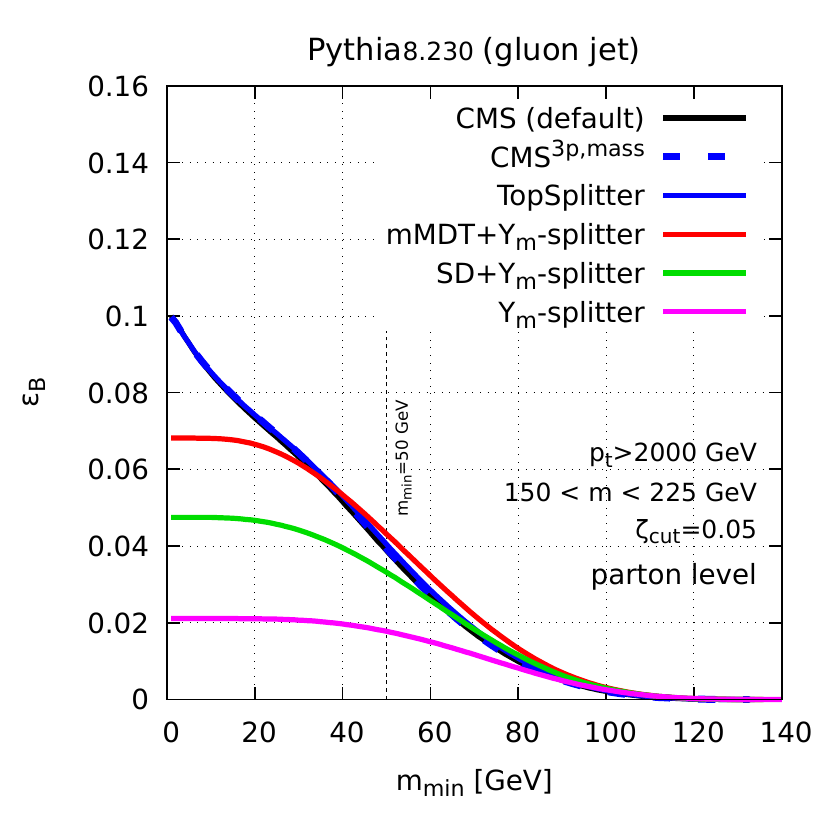}
  \hfill
  \includegraphics[width=0.5\textwidth]{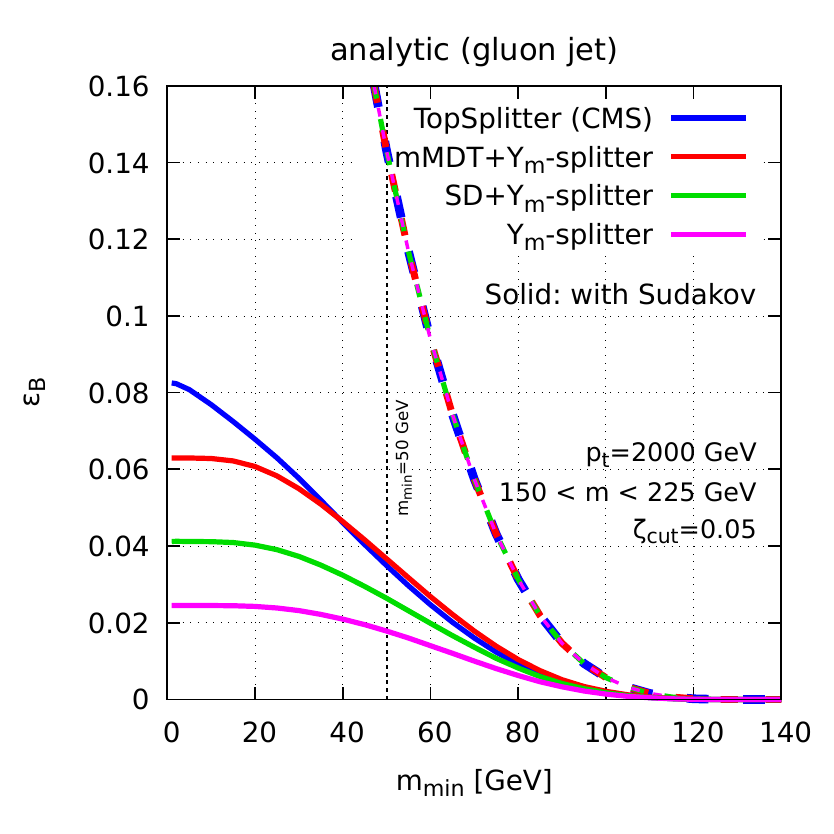}
  \caption{Same as Fig.~\ref{fig:mmin-distrib} but this time for
    gluon-initiated jets. Also shown in the analytic calculations on
    the right is the result without the application of the Sudakov
    form factors, using just the pre-factor from the triple-collinear
    splitting function for an initial gluon.}
  \label{fig:mmin-distrib-gluons}
\end{figure}

For completeness, we show the results obtained for gluon-initiated
jets in Fig.~\ref{fig:mmin-distrib-gluons}.
Again, the overall behaviour and ordering between the taggers are
correctly reproduced by the analytic calculations. However, we see
larger quantitative differences, in particular at small $m_\text{min}$
than what was seen for quark-initiated jets. This is likely due to the
fact that the Sudakov factors are larger in the case of gluon jets,
e.g.\ we see a suppression by a factor $\sim 4-6$ for
$m_\text{min}=50$~GeV, relative to the result without a Sudakov shown
using the dashed lines in
Fig.~\ref{fig:mmin-distrib-gluons}. Therefore, subleading corrections,
not included in our calculation, also have a larger impact.
Since the QCD background in the boosted limit is dominated by
quark-initiated jets ($>80$\% at 2~TeV), this has little impact on
practical applications (and we will focus on quark-initiated jets in
what follows).

Lastly, from the results of this section alone it may be tempting to
conclude that the \Ym-splitter method should be the preferred option
for top tagging. Indeed for studies of jet substructure with signal
jets initiated by a colourless electroweak boson decay, the impact on
the QCD background was most often the decisive factor in dictating
tagger performance \cite{Dasgupta:2013ihk}. In the present case with a
coloured parton also initiating the signal, one also needs to consider
the impact of QCD radiation for the signal jets too. The final word on
tagger performance will therefore involve also an analysis of the
signal, which is the subject of the next section.

\section{Signal efficiency and performance}
\label{sec:signalandroc}
Having studied the action of top taggers on QCD background jets we
shall here consider the case of signal jets. As a consequence we shall
then produce ROC curves purely from analytics and compare those to
curves obtained from Monte Carlo event generators. Finally we shall
examine here the role of non-perturbative effects.

\subsection{Signal efficiency}\label{sec:top-calculations}
To study signal jets we consider the case of boosted top production in
a given hard process, with the top exhibiting a three-pronged hadronic
decay to a $b$ quark and a hadronically decaying W boson, which form
the constituents of the top jet at leading order. One then has to take
account of the action of the top taggers on the three-pronged system.

The basic leading-order result can be obtained as for the QCD case 
\begin{equation}
\label{eq:topprefac}
\frac{d\sigma}{d\rho}  = \int d \Phi_3 \,  \left| \Mtop \right|^2 \delta
\left(\rho-\frac{s_{123}}{R^2 p_T^2} \right) \Theta^{\mathrm{tagger}}\left
(\rho_{\mathrm{min}},\zetacut \right) \, \Theta^{\mathrm{jet}},
\end{equation}
where $\Mtop$ is the matrix element for the top decay process,
$d\Phi_3$ the three-body phase space in the collinear limit as before,
and with the tagger and jet finding conditions now applied to the top
decay products. For an on-shell top quark we have that
$\rho = \rho_t = \frac{m_t^2}{R^2 p_T^2}$.  The above result is simple
to compute numerically and in implementing the result for the squared
matrix-element we have made the usual substitution of the W boson and
top quark propagators by a Breit-Wigner form.

In contrast to the case of colour singlet (e.g.\ W/Z/H) decays however,
the above tree-level result is not sufficient to give a good
description of substructure and tagging efficiency for top jets. The
obvious reason for this is that the top quark is a coloured object and
hence one must consider the role of accompanying QCD radiation. Soft
gluons are emitted by the virtual top quark in the course of producing
an on-shell final-state top and further emissions occur during the
top-decay process.

While multiple soft emissions are generally thought to be less
significant in heavy quark production than is the case for light
quarks, in the highly boosted region where $m_t^2 \ll p_t^2$ the top
quark can be considered as being essentially light and all-order
resummation effects start to become important.  In particular, in the
boosted regime, we may ignore the effect of the dead cone
\cite{Dokshitzer:1991fd} of order $m_t^2/p_t^2 \sim \rho$, which does
not play a role at the logarithmic accuracy in $\rho$ that we are
concerned with here.  At the same time while soft gluon emission in
top production and decay is known to have a complicated emission
pattern \cite{Dokshitzer:1992nh} especially for gluon energies near or
below the top width, again at our leading double-logarithmic accuracy
where we are concerned with only soft and relatively collinear
radiation, these complications can be neglected. Hence one can treat
the radiation as for the massless case to be essentially stemming from
a single fast moving colour line along the jet direction.

The soft emissions from the top quark, which are recombined into the
top jet, will contribute to the mass of the jet. We consider the jet
mass distribution after the further application of the various
top-taggers which, as for the case of the QCD background, places
constraints on the accompanying soft gluon emission within the top
jet, and leads to Sudakov form factors multiplying the top production
and decay probability.

Given our calculations in the previous sections for QCD jets it should
simple to understand the basic features of the resulting Sudakov
factor multiplying the leading-order electroweak factor
Eq.~\eqref{eq:topprefac}, for the different taggers. Two important
differences from the QCD case are firstly that the electroweak top
decay treated via the pre-factor already dominates the jet mass
condition since it produces a jet mass equal to the top mass for an
on-shell top decay, and secondly the fact that the W boson radiated
off the top is a colour singlet and hence does not radiate gluons
unlike a primary gluon emission from say a quark jet which, as we have
accounted for in the QCD jet case, acts as a source of relevant
secondary emissions.

A treatment of signal jets at the same level as we have performed for
background jets, i.e.\ one where $\ln \zetacut$ and
$\ln \rho/\rho_{\mathrm{min}}$ are also resummed, proves to be
substantially more complicated, e.g.\ because the ordering of the three
prongs found by the taggers' double declustering procedure will in
general involve different combinations of the $b$ quark and the $W$
decay products.
Additionally, gluon emissions from the top could contribute to shifting
its mass. Their effect would depend on both their interplay with the
tagger (including the dynamics of the three top decay products) and on
the mass window cut imposed on the tagged jet. The latter introduces
yet another non-trivial scale in the calculation.

As a consequence of these extra complications, for top jets we shall
not try to achieve a full double-logarithmic accuracy including
logarithms of $\rhomin/\rho$ and $\zetacut$ on equal footing with
logarithms of $\rho$.
Instead, we shall primarily focus on getting the dominant behaviour in
$\rho$.
Since we are also interested in investigating the strongly-ordered
limit, we will use a mass-like Sudakov down to the scale
$\text{min}(\rho_1,\rho_2)$. This means that \TopSplitter and
mMDT+\Ym-splitter would use a mMDT Sudakov $R_{\text{mMDT}}$,
Eq.~(\ref{eq:radiator-ysplit-mmdt-mmdt}), SD+\Ym-splitter would use a
Soft-Drop Sudakov $R_\text{SD}$, Eq.~(\ref{eq:RSD}), and ungroomed
\Ym-splitter would use a plain jet-mass Sudakov
$R^{\text{(primary)}}_{\text{\Ym-splitter}}$, all taken at the scale
$\text{min}(\rho_1,\rho_2)$.
In the case of the \Ym-splitter taggers, $\text{min}(\rho_1,\rho_2)$
is by definition equal to $\rho_2$ in the strongly-ordered limit and
is the natural scale for the Sudakov.
In the case of \TopSplitter, one could instead expect a mixture of the
$\theta_1^2$ angular scale and the $\rho_2$ mass-like scale
(cf. Fig.~\ref{fig:lund-cms}). Since one can trade $\theta_1^2$ for
$\rho_1$ up to subleading logarithms of $\zetacut$, the scale
$\text{min}(\rho_1,\rho_2)$ is also appropriate.

We have also investigated the impact of other choices which have the
same formal accuracy as the choice mentioned above. Specifically, we
have checked that the corrections, compared to using the same form of
the Sudakov, but taken at the scale $\rho$, were within 20\% in the
phenomenologically relevant region, which should really be seen as the
ballpark uncertainty on our calculations for signal
jets. Additionally, we have also considered using the full primary
Sudakov form factor, derived for quark jets in
Section~\ref{sec:analytic-resum}, which should also achieve the job of
capturing the bulk of the radiation from the top and bottom quarks in
the strongly-ordered limit. We found results very similar to the ones
obtained with the simpler mass-like Sudakov taken at the scale
$\text{min}(\rho_1,\rho_2)$ and hence we continue to use the latter as
our default form.

Our results for the top distribution can therefore be written as
\begin{equation}\label{eq:topfull}
\frac{\rho}{\sigma}\frac{d\sigma}{d\rho}
= \int d \Phi_3 \,  \left| \Mtop \right|^2
\rho \delta \left(\rho-\frac{s_{123}}{R^2 p_T^2} \right)
\Theta^{\mathrm{tagger}}\left(\rho_{\mathrm{min}},\zetacut \right)\,
\Theta^{\mathrm{jet}}\exp \left[-R_{\text{tagger}}^{\text{(mass)}} \right],
\end{equation}
with
\begin{align}
R_{\text{\TopSplitter}}^{\text{(mass)}} =
  R_{\text{mMDT+\Ym-splitter}}^{\text{(mass)}} & =
  R_{\text{mMDT}}(\min(\rho_1,\rho_2)),\\
  R_{\text{SD+\Ym-splitter}}^{\text{(mass)}} & =
  R_{\text{SD}}(\min(\rho_1,\rho_2)),\\
  R_{\text{\Ym-splitter}}^{\text{(mass)}} & =
  R^{\text{(primary)}}_{\text{\Ym-splitter}}(\min(\rho_1,\rho_2)),
\end{align}
where $\rho_1$ and $\rho_2$ are defined according to
Eq.~(\ref{eq:def-rhos-topsplitter}) for \TopSplitter and
Eq.~(\ref{eq:def-rhos-ysplitter}) for the \Ym-splitter
variants.

\begin{figure}
  \includegraphics[width=0.5\textwidth]{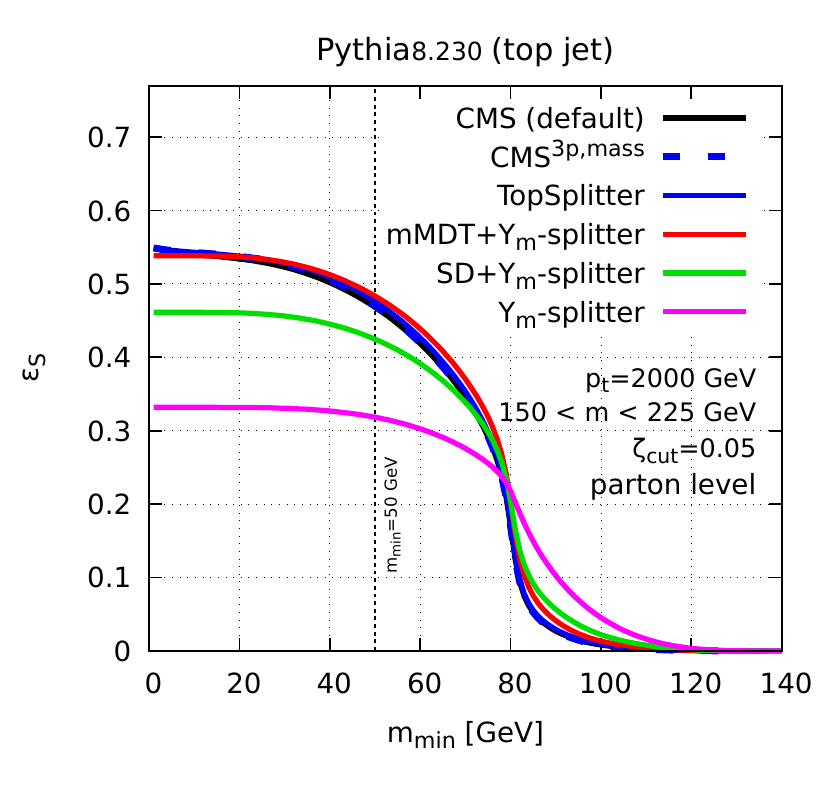}
  \hfill
  \includegraphics[width=0.5\textwidth,page=1]{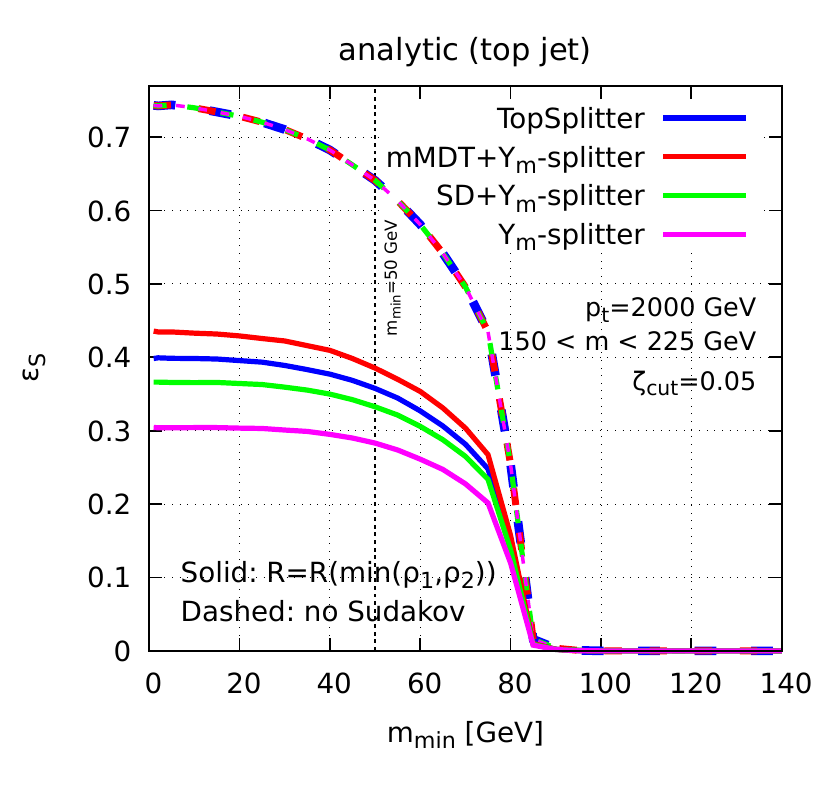}
  \caption{Same as Fig.~\ref{fig:mmin-distrib} this time for signal
    (top) jets. Also shown is the analytic result without the
    inclusion of a Sudakov form factor.}
   \label{fig:mmin-distrib-top}
\end{figure}

With the top-decay result supplemented by Sudakov form factors we
ought to be able to capture the main features seen in the performance
of taggers using Monte Carlo event generators.
To that aim, we show the signal efficiency as a function of
$m_\text{min}$ in Fig.~\ref{fig:mmin-distrib-top}.
We see that the effects of the Sudakov seem over-estimated in the
analytics relative to the MC results, but that ordering between the
taggers is reasonably reproduced. We also show in
Fig.~\ref{fig:mmin-distrib-top}, via the dashed curves, the impact of
not including any Sudakov form factor in the analytic calculations for
signal jets, in which case the analytic results are the same for all
taggers and differ substantially from Pythia results.

The analytical result shows also a minor difference between
\TopSplitter and mMDT+\Ym-splitter, where our analytic calculation
predicts a larger suppression for the latter which is not observed in
the Pythia simulations.
Since our treatment of top jets does not reach the same accuracy as
what was obtained for QCD jets in Section~\ref{sec:analytic-resum},
and since the MC simulations themselves do not contain information for
example on the triple-collinear phase space, such differences between
analytics and MC results should be expected. In this precise case of
comparing \TopSplitter with mMDT+\Ym-splitter, the observed difference
has to be driven by the different definitions for $\rho_1$ and
$\rho_2$ (as a function of the parton kinematics from the
triple-collinear splitting), Eqs.~(\ref{eq:radiator-ysplit-mmdt})
and~(\ref{eq:def-rhos-topsplitter}).
Indeed, while for QCD jets we expect emissions with momentum fractions
close to $\zetacut$, the situation will be more symmetric for top
jets, meaning in practice a smaller value for $\rho_1$ and $\rho_2$ in
the case of \TopSplitter compared to mMDT+\Ym-splitter. This
smaller value translates in a larger
$R_{\text{mMDT}}(\min(\rho_1,\rho_2))$ and hence a smaller signal
efficiency for \TopSplitter (again, compared to
mMDT+\Ym-splitter).\footnote{If we were instead using a simple mass
  Sudakov taken at the scale $\rho$ for top jets --- achieving the
  same formal accuracy as what have used so far --- we would obtain
  the exact same signal efficiency for \TopSplitter and
  mMDT+\Ym-splitter.}
These differences are clearly beyond our targeted accuracy.

The main message that emerges from our studies in the current section
is that a Sudakov form factor is essential to describe the behaviour
of the taggers on signal jets. The basic form of the Sudakov that we
have used in the signal case is sufficient to understand the main
features of top taggers but a more precise statement on tagging
efficiency, as we have for instance for QCD background jets, would
require a more detailed analytic calculation for signal jets which is
beyond the scope of our present work.
Finally we remark that on the Monte-Carlo side, we also note that no
observable differences are seen in Fig.~\ref{fig:mmin-distrib-top}
between the various CMS-related taggers.
In the following section we will look at tagger performance using
both parton shower and analytic methods.

\subsection{Performance and non-perturbative effects}\label{sec:top-performance}

We now discuss the performance of the various taggers using the
standard ROC curves which show the background efficiency or mistag
rate plotted against the signal efficiency. For a given signal
efficiency, the tagger with the lowest mistag rate is considered the
most performant.

A point that is worth noting is that due to a very similar Sudakov
suppression seen for the signal and the background, any gains that are
produced by Sudakov suppression of emissions from a QCD jet are
largely offset by a corresponding suppression of the signal.
Therefore a large Sudakov suppression is not necessarily beneficial
for the case of top tagging in contrast to the tagging say of colour
singlet electroweak and Higgs bosons.  An exception to the above may
in principle be expected to occur for the case of gluon jets where the
Sudakov suppression of the background is indeed more than that for the
signal, owing to the larger colour factor for emissions from gluon
jets. In general however the background will be a mix of quark and
gluon jets, with the quark jet component being dominant at higher
$p_t$ where Sudakov effects are stronger for a fixed jet mass.  For
this reason we start by looking at the highest phenomenologically
relevant $p_t$ values, i.e.\ in the TeV region, with quark jets alone.

\begin{figure}
  \includegraphics[width=0.48\textwidth]{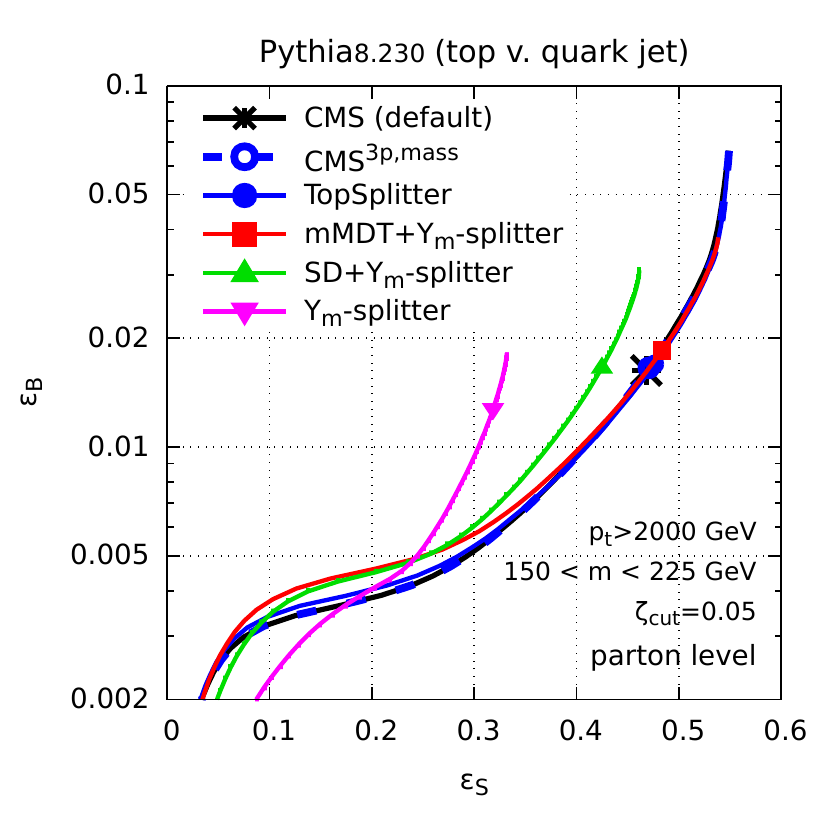}
  \hfill
  \includegraphics[width=0.48\textwidth]{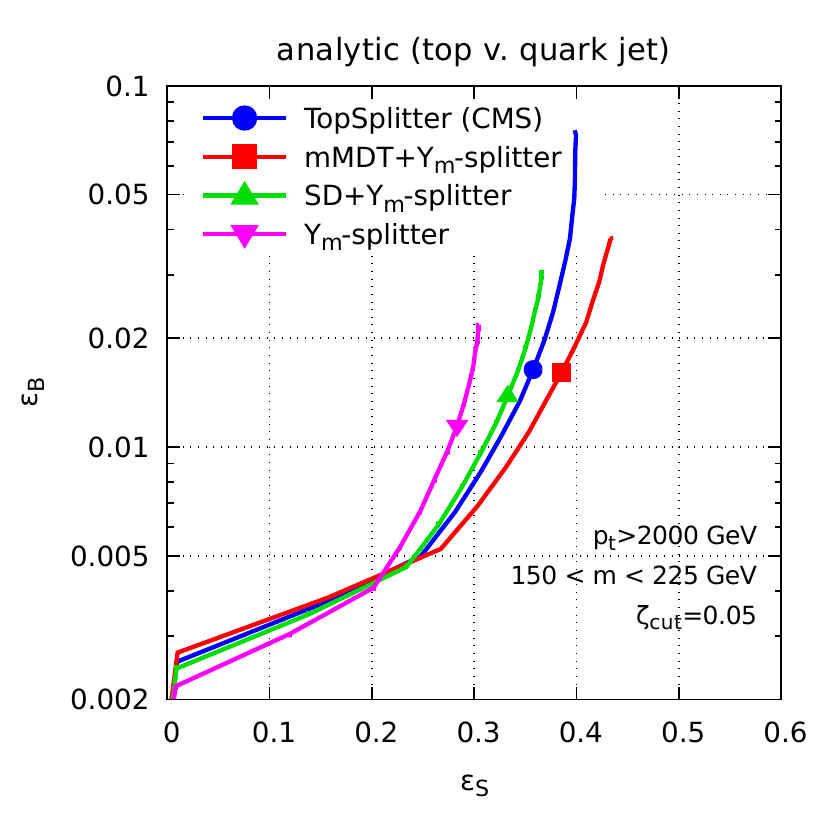}
  \caption{ROC curves corresponding to the $m_{\text{min}}$ scan shown
    in Fig.~\ref{fig:mmin-distrib}.
    This time, the Pythia results also include the default CMS tagger for
    comparison (in black).
    The thicker points correspond to the default value
    $m_{\text{min}}=50$~GeV.  }\label{fig:rocs}
\end{figure}

Figure \ref{fig:rocs} shows the ROC curves one obtains for
$p_t = 2 \, \mathrm{TeV}$ with a pure quark jet background.
The curves correspond to tagging in a mass window $150<m<225$~GeV, use
$\zetacut = 0.05$ , as done throughout our studies, and both parton level
results from Pythia (left) as well as analytical results (right) are
shown. A first observation is that except at fairly low signal
efficiencies, a larger Sudakov results in a larger mistag rate for a
given signal efficiency i.e.\ a worse performance. Based on this
observation we would expect to see a definite ordering in the results
for tagger performance. From an analytical viewpoint the smallest
Sudakov suppression belongs to \TopSplitter (and
$\mathrm{CMS^{3p,mass}}$) and mMDT+\Ym-splitter taggers.
A somewhat larger Sudakov suppression is seen in our analytic
calculations for SoftDrop~($\beta=2$) and the largest suppression is
for \Ym-splitter with a double-logarithmic plain-mass type
Sudakov. 
This globally corresponds to the ordering seen in the analytic ROC
curves above a signal efficiency $\epsilon_s \sim 0.2$. Instead for
lower signal efficiencies the ordering is inverted so that taggers
with a large Sudakov perform better. A larger signal efficiency is
however what we clearly would desire from a phenomenological viewpoint
and so taggers with a smaller Sudakov would be favoured.  The results
from the Pythia parton shower are in general agreement with our
analytical expectations and a similar ordering is seen for those
results. However, while the analytic results indicate some difference
in performance between \TopSplitter and mMDT+\Ym-splitter, these are
seen to perform essentially identically in Monte Carlo studies at
higher signal efficiencies. Such differences can be easily ascribed to
the less precise treatment of the signal Sudakov in the analytics,
discussed in Section~\ref{sec:top-performance}.

It is also noteworthy that no differences are seen in parton shower
results between the default CMS tagger, \TopSplitter,
$\mathrm{CMS^{\mathrm{3p,mass}}}$ and the mMDT+\Ym-splitter methods,
at least for efficiencies $\epsilon\gtrsim 0.35$.
Apart from the IRC unsafe CMS tagger all these other methods have the
common feature of an essentially mMDT style Sudakov at low masses,
albeit with differences of detail.  The main message appears to be
that it is possible to create a family of taggers which are IRC safe
but give a similar performance to the default CMS tagger with the
family being defined by its key feature of an mMDT style Sudakov.

Finally, we show in Appendix~\ref{sec:perf-low-pt} that our observations
are still valid at lower jet $p_t$ (1~TeV or even lower down to about
500~GeV) albeit with a reduced difference between the taggers,
attributed to a reduction of the phase-space available for radiation
and the decreased importance of Sudakov effects. In particular, it
means that the \TopSplitter can be considered as an effective and more
robust replacement of the CMS top tagger over a wide range of $p_t$
values relevant to phenomenology.

A discussion of tagger performance and reliability is not complete
without a discussion of non-perturbative effects. As we mentioned
before, ROC curves produced using event generators are subject to a
theoretical uncertainty. However estimating the uncertainty on such
ROC curves is a far from simple exercise even conceptually, largely
owing to the sole reliance on Monte Carlo event generators. It is
however safe to say that results for methods which are either IRC
unsafe like the CMS tagger, or those that receive large
non-perturbative corrections, must be considered to suffer from a
larger theoretical uncertainty than IRC safe methods which
additionally show only small non-perturbative corrections, even if
that uncertainty cannot be easily quantified. Therefore examining the
impact of non-perturbative corrections is important in order to more
reliably assess the performance of a tagger.

\begin{figure}
  \includegraphics[width=0.48\textwidth,page=1]{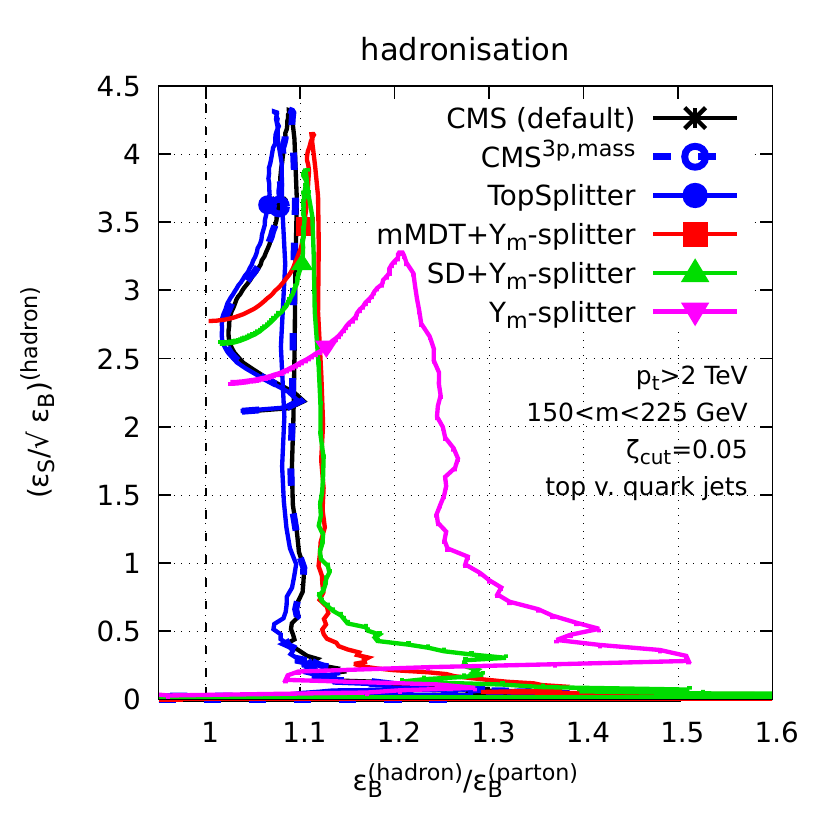}
  \hfill
  \includegraphics[width=0.48\textwidth,page=2]{figs/np-effects.pdf}
  \caption{Both plots show how the sensitivity to non-perturbative
    effect ($x$ axis) and the discriminating power ($y$ axis) evolve
    when varying the cut on $m_{\text{min}}$ for different taggers.
    Left: effects observed when switching on hadronisation, i.e.\ going
    from parton level to hadron level.
    Right: effects observed when including the Underlying Event.
    The symbols correspond to $m_{\text{min}}=50$~GeV.}\label{fig:np-effects}
\end{figure}

Figure \ref{fig:np-effects} shows a plot of the signal efficiency
divided by the square-root of the background efficiency, also known as
the signal significance, which quantifies the tagger performance on
the $y$ axis, while at the same time showing the sensitivity to
non-perturbative effects on the $x$ axis. To estimate the latter, the
measure we have chosen is the ratio of the background efficiency at
hadron level to that at parton level to assess the impact of
hadronisation (for a fixed $m_\text{min}$ cut) in the left plot, and
the ratio of the background efficiency at hadron level including UE to
that without the UE on the right plot of Figure
\ref{fig:np-effects}. Similar studies have also been carried out in
the past for the case of W/Z/H tagging, for instance in
Ref.~\cite{Dasgupta:2016ktv}.

A number of points follow from consideration of Figure
\ref{fig:np-effects}. Firstly the inclusion of non-perturbative
effects as measured by the deviation of the results from unity along
the $x$ axis does not have a very substantial effect for a wide range
of signal significances, with the notable exception of \Ym-splitter
which due to its inherent lack of grooming suffers significantly from
both hadronisation and UE effects. For other methods the hadronisation
effects are no larger than around the $15 \%$ level with even smaller
effects for the default CMS tagger, $\mathrm{CMS^{3p,mass}}$ and
\TopSplitter. A remarkable degree of similarity and effectiveness
across methods is seen with regard to removal of contamination from
the UE with the only exception again being \Ym-splitter which is
entirely expected from previous studies of Y-splitter
\cite{Dasgupta:2015yua} . On this basis the non-perturbative studies
do not have any sizeable impact on the main conclusions that we
reached from the parton level analysis before. The best taggers in
terms of sheer performance are also the ones which are most resilient
to non-perturbative effects, which is in contrast to what is seen for
W/Z/H tagging where \Ym-splitter followed by grooming using mMDT or
trimming far outperforms other methods, but at the cost of large
non-perturbative effects. When one factors in IRC safety which is a
key element in assessing the robustness of a tool, one should replace
the CMS tagger with either $\mathrm{CMS^{3p,mass}}$ or \TopSplitter
which leads to no loss of performance. If one further factors in
analytic calculability then the \TopSplitter method emerges as the one
over which we have the best theoretical control certainly for
phenomenologically relevant $m_{\mathrm{min}}$ values, while at the
same time maximising the performance,

Finally, we have also tested that these conclusions remain valid down
to (at least) jet $p_t$'s of 500~GeV, where the \TopSplitter
non-perturbative corrections remain in the 15-20\% range, followed by
mMDT+\Ym-splitter and SD+\Ym-splitter around 30\%.
In this context, it would also be interesting to investigate a version of
the \Ym-splitter tagger where one first applies Recursive Soft
Drop~\cite{Dreyer:2018tjj}, e.g.\ with two layers of grooming with
$\beta=0$ (``recursive mMDT'') or $\beta=2$, or infinite recursion
with $\beta=2$.

\section{Conclusions}
\label{sec:conclusions}
In this article we have studied aspects of top-tagging from first
principles of QCD using the methods of analytic resummation supported
by Monte Carlo studies. The aim has been to try and identify the main
physical principles that are at play and hence better understand the
effect of using top tagging methods on background and signal jets.

To consider an explicit example of a tool that has been used directly
in the context of LHC phenomenology, we started by studying the CMS
top tagger.  Here we discovered a potentially serious flaw, namely
that of collinear unsafety at high $p_t$.  The collinear unsafety of
the CMS tagger was seen to originate in the step of selecting three
prongs from four on the basis of their energy. Hence we proposed
variants of the CMS tagger that are explicitly IRC safe even at high
$p_t$. One variant that we named $\mathrm{CMS^{3p,mass}}$ selects
three prongs from four based on the invariant mass while another
variant we named \TopSplitter, selects the emission which dominates
the prong mass in the soft limit as a product of the
declustering. While both methods are collinear safe, \TopSplitter is
simpler from the viewpoint of the analytical calculations we aimed at
in this article.

In addition to the above methods which are all based on C-A
declustering of a jet, we introduced new methods based on gen-$k_t$
declustering. Here we adapted our previously suggested \Ym-splitter
method \cite{Dasgupta:2016ktv} for use in top tagging. Our earlier
studies based on W/Z/H tagging have shown that \Ym-splitter when
additionally supplemented by some form of grooming has the potential
to be a high performance tool \cite{Dasgupta:2016ktv}, which led us in
this paper to investigate a combination of grooming with \Ym-splitter.

For the QCD background, we carried out leading logarithmic in jet mass
analytical calculations for \Ym-splitter, mMDT + \Ym-splitter,
SoftDrop ($\beta=2$) + \Ym-splitter, \TopSplitter and the
$\mathrm{CMS^{3p,mass}}$ taggers. For all but the last case we were
able to supplement a resummation of large logarithms in the jet masses
$\rho$ or $\rho_{\mathrm{min}}$ with additional resummation of leading
logarithms in $\zetacut$ and $\rho/\rho_{\mathrm{min}}$, counting them
on the same footing as logarithms of $\rho$ or $\rho_{\mathrm{min}}$.
Our results were seen to take the form of an order $\alpha_s^2$
pre-factor which multiplies a Sudakov exponent arising from
resummation. We argued that an accurate calculation of the pre-factor
should require going beyond the picture of strong ordering in angles
or energies of emissions and should involve instead the use of
triple-collinear splitting functions.  Such splitting functions are
not included in the Pythia shower, or indeed in other well-known
showers, commonly used to study tagger performance. Ultimately however
the triple-collinear splitting functions gave a somewhat modest
$\sim10 \%$ effect for $m_{\mathrm{min}} =50 \, \mathrm{GeV} $
relative to the strong angular-ordering approximation which is in
principle correctly included in the Pythia shower.

A comparison of our analytical calculations for QCD background jets
with the Pythia parton shower revealed general good agreement across a
wide range of $m_{\mathrm{min}}$ values and excellent agreement at the
phenomenological working point of
$m_{\mathrm{min}}=50 \, \mathrm{GeV}$ for all methods for which
resummed results exist (i.e.\ all our taggers except the
collinear-unsafe default CMS tagger.)
Our full resummation including logarithms of $\zetacut$ and
$\rho/\rho_{\mathrm{min}}$ was seen to be required in order to obtain
better agreement with Pythia and becomes crucial to include especially
at small $\rho_{\mathrm{min}}$. The basic conclusion from our analytic
versus Monte Carlo comparisons is that we appear to have very good
analytic control over top taggers studied and developed in this paper,
when applied to QCD jets

In terms of performance we have found that, as may readily be
anticipated, taggers with a larger Sudakov suppression are more
effective at removing the QCD background. Our analytics suggest that
\Ym-splitter with its plain jet mass type Sudakov suppression should
therefore produce the lowest background mistag rate and this
expectation is confirmed by the Pythia shower. We also found that the
ordering of background mistag rates between taggers which emerges in
our analytics is indeed reproduced in the Pythia shower at parton
level. It is noteworthy that the default CMS tagger produces identical
results in the Pythia parton shower to our newly-proposed alternatives
$\mathrm{CMS^{3p,mass}}$ and \TopSplitter.

We also studied the effect of top taggers on signal jets initiated by
a top quark.  The resulting jet mass distributions also receive a
Sudakov suppression factor similar to that for QCD background, due to
the colour charge of the top quark, although here our analytical
calculations were less precise than those we carried out for the QCD
background and we neglected retaining full control over logarithms of
$\zetacut$ and $\rho/\rho_{\mathrm{min}}$.
We discovered that the impact on signal together with background is
such that, at high $p_t$, taggers with a larger Sudakov suppression
generally perform less well, at least for reasonably large signal
efficiencies, than those with a smaller Sudakov, assuming a pure quark
background. Therefore the plain \Ym-splitter method is less performant
than \Ym-splitter with SoftDrop ($\beta=2$) pre-grooming, in turn less
performant than \Ym-splitter with mMDT pre-grooming which produces a
Sudakov which resembles more closely the mMDT Sudakov, rather than the
plain mass type of Sudakov seen with \Ym-splitter.  The mMDT
pre-groomed \Ym-splitter, \TopSplitter, the default CMS tagger and the
$\mathrm{CMS^{3p,mass}}$ tagger gave essentially identical performance
at signal efficiencies larger than about $0.35$, i.e.\ the Pythia ROC
curves for these methods coincide. For lower signal efficiencies the
default CMS tagger, $\mathrm{CMS^{3p,mass}}$ and \TopSplitter still
remain very close to one another in performance. The analytics however
suggested some modest differences also between \TopSplitter and
mMDT+\Ym-splitter even at higher signal efficiencies, which was not
seen in the parton shower studies.

We evaluated also the role of non-perturbative effects. With the
exception of plain \Ym-splitter we noticed that all the taggers are
quite resilient to non-perturbative effects due to their inherent
grooming aspect. Hadronisation effects were found to be no more than
$\sim 15 \%$ for phenomenologically relevant values of signal
efficiency while the underlying event contribution was generally less
than a few percent.

Overall we found that it is possible to develop a range of IRC safe
methods, for tagging three-pronged jet substructure, which can be
understood from first principles of QCD (i.e.\ largely independently
from MC results). The more performant techniques at high $p_t$ are
ones which have a common feature of a smaller mMDT style Sudakov
suppression. Our alternatives to the default CMS tagger are virtually
identical in performance to the CMS tagger, with \TopSplitter emerging
as our preferred method, due to the higher accuracy of the
corresponding analytical calculation.

Most importantly, armed with a range of methods and a detailed
understanding of their impact, we have acquired both some flexibility
and insight which will be important for also studying the optimal
combination of top taggers with jet shape variables such as
$N$-subjettiness or energy correlation functions, and to explore the
origin and nature of the further gains due to using jet shapes. In
future work we intend to enhance our understanding of top-tagging by
considering such combinations, which are widely used in LHC studies,
also from an analytical viewpoint.

\section*{Acknowledgements}
MD thanks the U.K's STFC for financial support via grant
ST/P000274/1. MD also thanks the CERN theoretical physics department
for a scientific associateship and for hospitality during the course
of this work as well as the School of Physics and Astronomy at the
University of Manchester for sabbatical leave which facilitated this
work. MD thanks the CEA Saclay and the French CNRS for financial
support and hospitality during the course of this work.  JR thanks the
U.K's STFC for financial support via grant ST/N504178/1. GS is
supported in part by the French Agence Nationale de la Recherche,
under grant ANR-15-CE31-0016.

\appendix

\section{Collinear unsafety of the CMS tagger with no
    $\Delta R$ cut}\label{app:event2}

The collinear unsafety of the CMS top tagger can be explicitly shown
using a fixed-order study.
As described in section~\ref{sec:cms-definition}, the collinear
unsafety appears when some substructure can be found in both primary
prongs. This requires at least 4 particles in the jet.
One method to obtain such jets is to generate $e^+e^-$ collisions with
QCD particles in the final state and to boost the whole event along
the $x$ axis to obtain a collimated jet. 
In practice, we have used the
Event2~\cite{Catani:1996jh,Catani:1996vz} generator with a
centre-of-mass energy of 80~GeV, boosted to 1 TeV.
We then reconstruct the jets with the
Cambridge/Aachen~\cite{Dokshitzer:1997in} with $R=1$ and keep jets
above 500~GeV. We measure the cross-section for the jets to pass either the
CMS top tagger or the CMS$^{3p,\text{mass}}$ with $\zetacut=0.05$ and
$m_\text{min}=30$~GeV.

This setup allows us to study the boosted jets to order $\alpha_s$
(with up to 3 particles in the jet) and $\alpha_s^2$ (with up to 4
particles in the jet). We note that since the tagger requires at least
3 particles in the jet, the order $\alpha_s$ is actually the leading
order here and there is no contribution from the 2-loop contribution
at order $\alpha_s^2$, which is not available in Event2.

\begin{figure}
  \centering
  \includegraphics[width=0.5\textwidth]{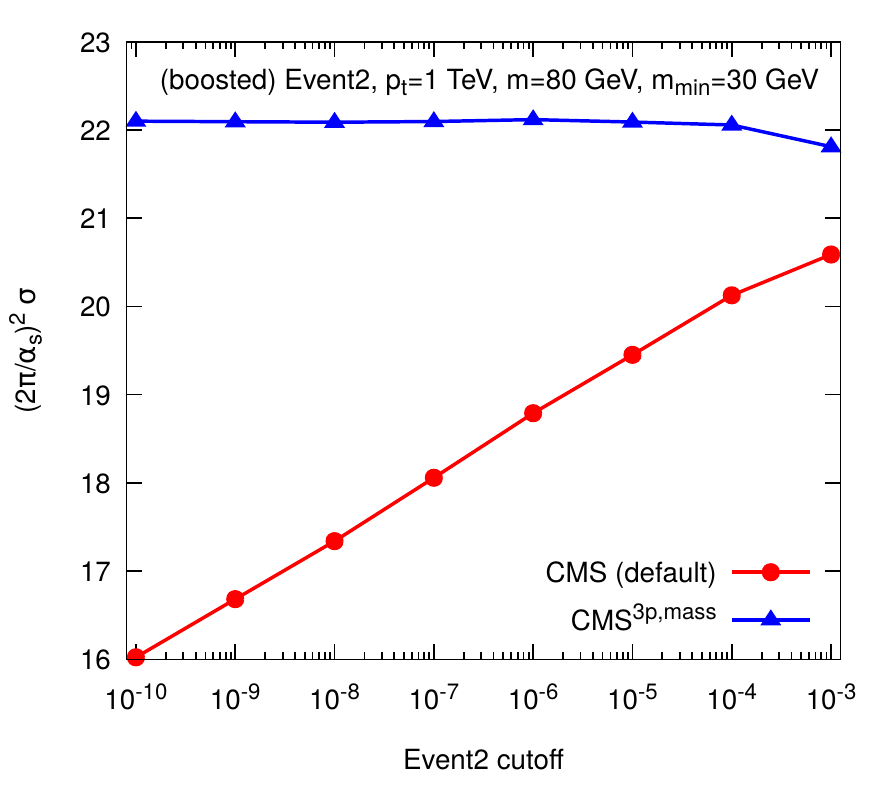}
  \caption{Cross-section for passing the CMS tagger as a function of
    the Event2 cut-off.}\label{fig:cms-coll-unsafety}
\end{figure}

In Fig.~\ref{fig:cms-coll-unsafety} we plot the cross-section for jets
passing the CMS tagger as a function of the internal cut-off used in
Event2.
For the default CMS top tagger, we see an obvious logarithmic
dependence on the cut-off as a result of the collinear unsafety of the
tagger.
Switching instead to the $\mathrm{CMS^{3p,\mathrm{mass}}}$ tagger, the
cross-section converges rapidly when the cut-off is decreased, showing
that the collinear unsafety has been cured.

We note that in the context of a resummed calculation, this collinear
unsafety will be tamed by the associated Sudakov form factor, i.e.\ the
default CMS tagger although collinear unsafe, remains Sudakov
safe~\cite{Larkoski:2013paa,Larkoski:2015lea}.
This potentially explains why little differences are seen in practice
between the CMS, CMS$^{\text{3p,mass}}$ and \TopSplitter taggers in
full Monte-Carlo simulations.
The collinear unsafety would however make it delicate to reliably
estimate the theoretical uncertainties associated with the CMS top
tagger.

\section{Variants of the CMS and Y-splitter taggers}\label{app:variants}

Here, we consider additional variants of the CMS
and \Ym-splitter taggers. We first define them and
then briefly compare them to the default versions discussed in the
main text.

\subsection{Definition of the variants}

The variants are as follows:
\begin{enumerate}
\item {\it $\zcut$ condition}: one can modify the CMS tagger such that one
  uses a $z_{\mathrm{cut}}$ type condition in performing the
  decomposition. This would involve a cut of the form
  $\frac{\mathrm{min}\left(p_{T,i},p_{T,j}\right)}{p_{T,i}+p_{T,j}}
  >z$ which uses the local $p_T$ of the cluster being decomposed,
  i.e.\ $p_{Ti}+p_{Tj}$ instead of the global $p_T$ of the hard jet in
  the denominator as is the case for the $\zetacut$ condition used
  originally by CMS and in the main body of the paper.
\item {\it $\rhomin$ condition only on secondary declustering}:
  variants where the taggers proceed exactly like the default
  $\mathrm{CMS^{3p,mass}}$, \TopSplitter and \Ym-splitter but we
  impose the $m_{\mathrm{min}}$ condition only on the 2 prongs
  produced in the secondary declustering instead of all 3 pairwise
  combinations.
\end{enumerate}

\subsection{Declustering with a $\zetacut$ or
  $z_{\text{cut}}$ condition}\label{app:local}
  
\begin{figure}
  \includegraphics[width=0.48\textwidth]{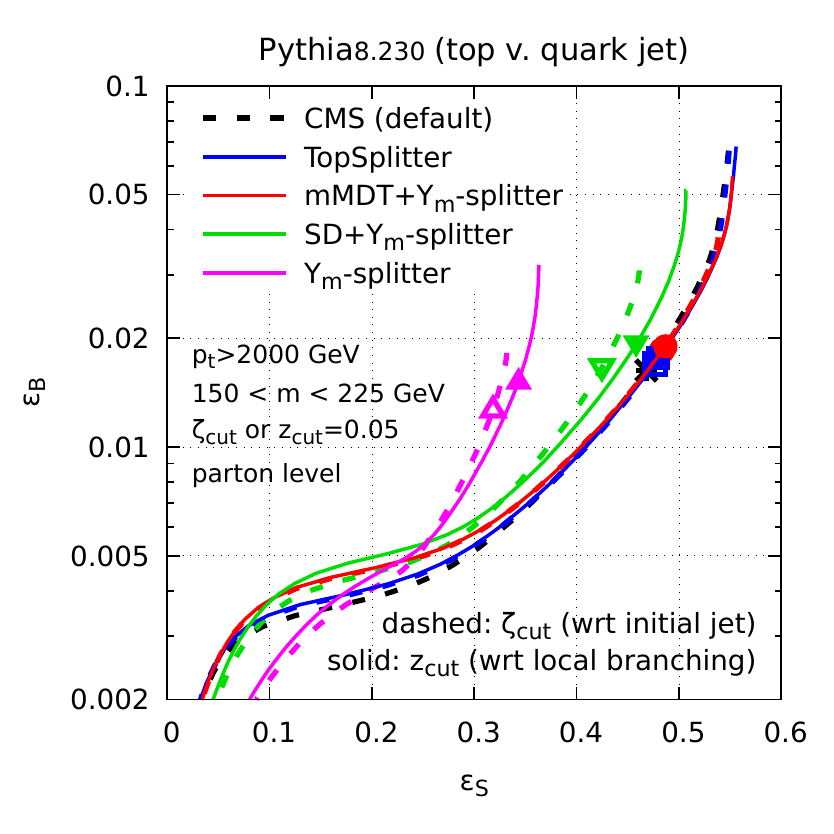}
  \hfill
  \includegraphics[width=0.48\textwidth]{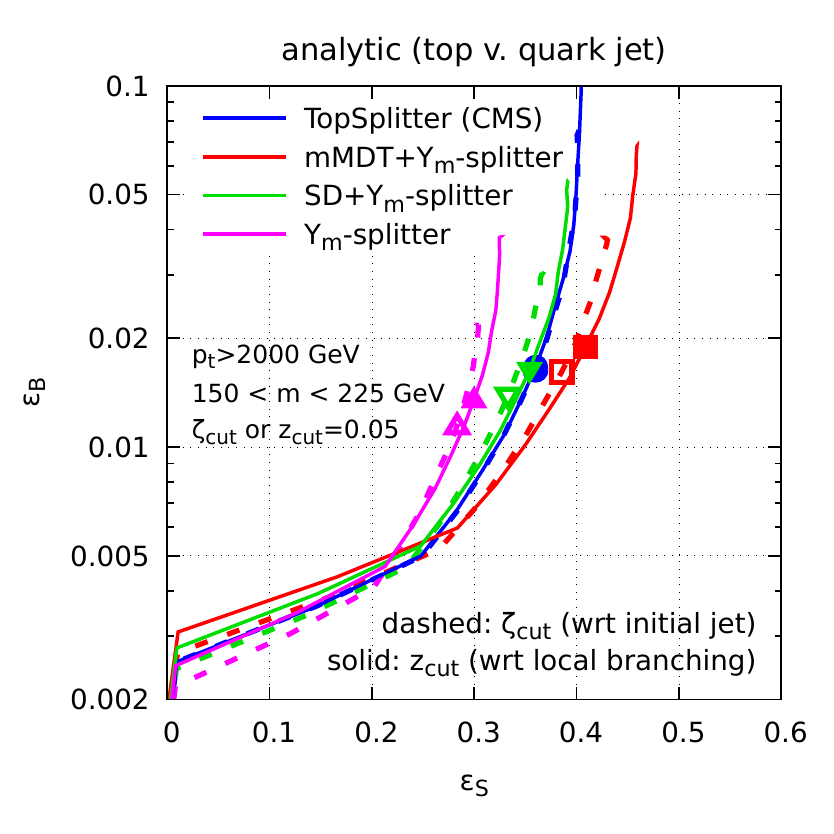}
  \caption{Comparison of the taggers performance when using a $\zcut$
    condition (solid lines) compared to the default $\zetacut$
    condition (dashed lines). Left: Pythia simulations, right: our analytic
    calculation.}\label{fig:rocs-zcut}
\end{figure}

We start by comparing the performance of the taggers when imposing a
$\zcut$ condition (compared to the default $\zetacut$ condition).
Our results are plotted in Fig.~\ref{fig:rocs-zcut} for Pythia
simulations (left plot) and for our analytic calculation (right plot).

Overall, we see little differences between the two variants, in
particular, for the CMS-related taggers.
For the \Ym-splitter taggers, we see a small difference in
performance, with the versions using a $\zetacut$ condition performing
marginally better at small signal efficiencies and the versions using
a $\zcut$ condition performing slightly better at large signal
efficiency.
Our analytic calculations reproduce these differences correctly
although the predicted difference in the case of mMDT+\Ym-splitter is
not seen in the Pythia simulations. This difference seems driven by
the signal (top) efficiency which is anyway not as well controlled
as the QCD background in our analytic calculations.

\subsection{Minimum pairwise condition v. secondary declustering condition}\label{app:1pair}
  
\begin{figure}
  \includegraphics[width=0.48\textwidth]{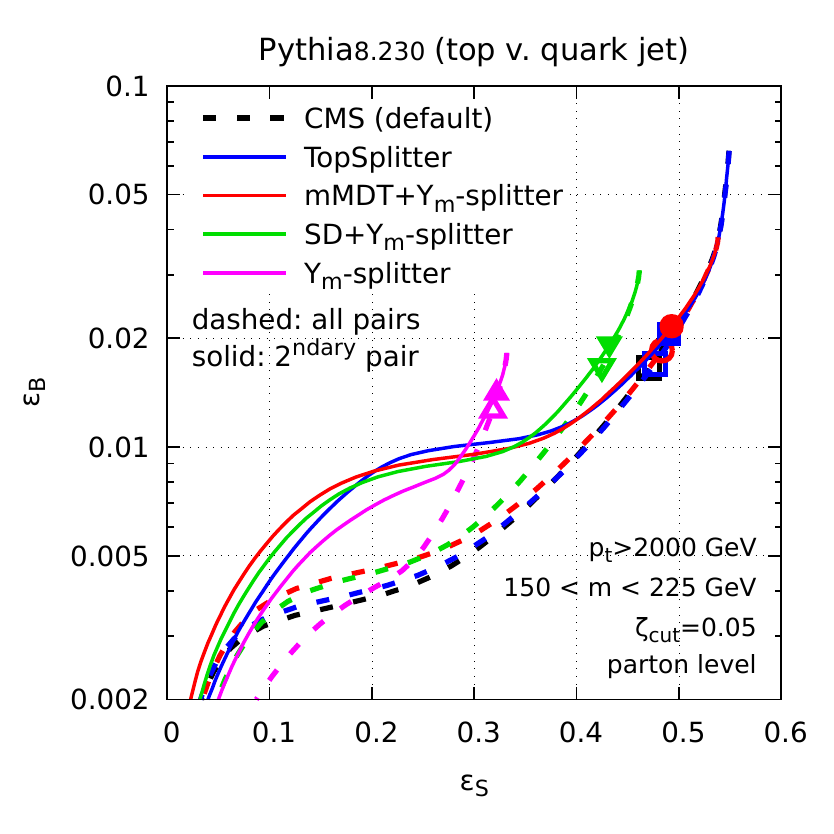}
  \hfill
  \includegraphics[width=0.48\textwidth]{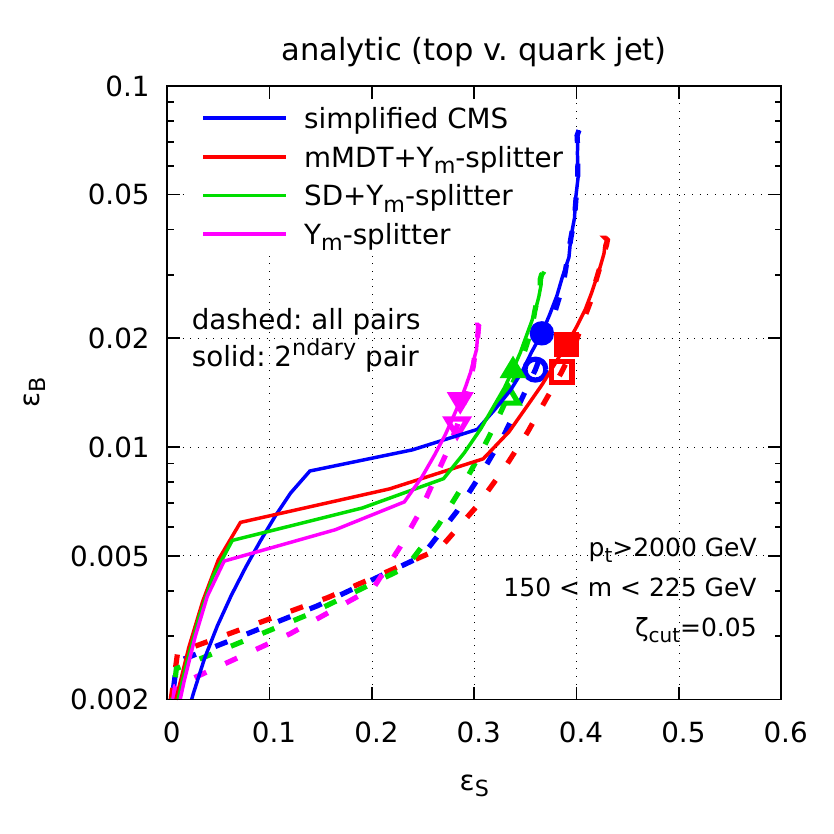}
  \caption{Comparison of the taggers performance when imposing the
    $\rhomin$ condition only on the secondary declustering (solid
    lines) compared to imposing the $\rhomin$ condition on all three
    pairwise masses (dashed lines). Left: Pythia simulations, right:
    our analytic calculation.}\label{fig:rocs-1pair}
\end{figure}

In Fig.~\ref{fig:rocs-1pair}, we compare the performance of the
variants of the taggers derived by imposing the $\rhomin$ condition
only on the secondary declustered branch, to the default \TopSplitter
and \Ym-splitter.
We see little difference between the default version (dashed lines)
and the corresponding variant (solid lines) at large signal
efficiency. However, at small signal efficiency, the default version
of the taggers clearly outperforms the variants, i.e.\ favouring the
case where the $\rhomin$ condition is imposed on the minimum pairwise
mass.
These behaviours are well captured by our analytic results.

\section{Analytic expressions for the radiators}\label{sec:radiators}

For completeness, we list in this appendix our results for the
radiators derived in section~\ref{sec:analytic-resum}, including
running-coupling effects and hard-collinear splittings.
Our results are written in terms of the ``building blocks'' introduced
in~\cite{Dasgupta:2015lxh}. For our purpose in this paper, the
only building block we need is\footnote{Compared
  to~\cite{Dasgupta:2015lxh}, we neglected the $\beta_1$ and $K$
  contributions, and we will introduce the ``$B$'' terms ---
  corresponding to hard collinear splittings --- via a shift of the
  $k_\text{max}$ argument.}
\begin{align}
  & T_{-\beta,2}(\kappa_\text{min},\kappa_\text{max};C_R)
  =  \int_0^1 \frac{d\theta^2}{\theta^2}\frac{dz}{z}
    \frac{\alpha_s(z\theta p_tR)}{2\pi}
    \Theta(z<\kappa_\text{max}\theta^\beta)\Theta(z\theta^2>\kappa_\text{min})\\
  &\quad = \frac{C_R}{2\pi\alpha_s\beta_0^2}\Big[\frac{U(\lambda_\text{max})}{1+\beta}+U(\lambda_\text{min})-\frac{2+\beta}{1+\beta}U\Big(\frac{\lambda_\text{max}+(1+\beta)\lambda_\text{min}}{2+\beta}\Big)\Big]\Theta(\kappa_\text{max}>\kappa_\text{min})\nonumber
\end{align}
with $\alpha_s\equiv\alpha_s(p_tR)$,
$\lambda_i=2\alpha_s\beta_0\ln1/\kappa_i$ and
$U(\lambda)=(1-\lambda)\ln(1-\lambda)$.
In particular, we have
\begin{equation}
T_{02}(\kappa_\text{min}, \kappa_\text{max};C_R)
  =  \frac{C_R}{2\pi\alpha_s\beta_0^2}\Big[U(\lambda_\text{max})+U(\lambda_\text{min})-2U\Big(\frac{\lambda_\text{max}+\lambda_\text{min}}{2}\Big)\Big]\Theta(\kappa_\text{max}>\kappa_\text{min})
\end{equation}
which corresponds to the standard mass Sudakov. We also note that
$T_{-\beta,0}$ vanishes in the $\beta\to\infty$ limit.
In practice, we have used a one-loop running coupling with
$\alpha_s(M_Z)=0.1383$ (matching the value used in Pythia).

With this at hand, we can write all the radiators introduced in
section~\ref{sec:analytic-resum} as follows:
\begin{align}
  R_{\text{\Ym-splitter}}^{\text{(primary)}}
  & = T_{02}(\rho_2,b_i;C_R),\\
  R_{\text{\Ym-splitter}}^{\text{(secondary)}}
  & = T_{02}(\rho_2/\theta_1,\rho_1/\theta_1b_g;C_A),\\
  R_{\text{SD+\Ym-splitter}}^{\text{(primary)}}
  & = T_{02}(\rho_2,b_i;C_R)-T_{-\beta,2}(\rho_2,\zetacut;C_R)
    + T_{02}(\rho_2/\theta_1,\zetacut\theta_1^{1+\beta};C_R),\\
  R_{\text{\TopSplitter}}^{\text{(red)}}
  & = T_{02}(\rho_2,b_i;C_R)-T_{02}(\rho_2,\zetacut;C_R),\\
  R_{\text{\TopSplitter}}^{\text{(blue)}}
  & = T_{02}(\zetacut\theta_1,\rho_2/\theta_1;C_R)
    - T_{02}(\zetacut,\rho_2;C_R),\\
  R_{\text{\TopSplitter}}^{\text{(secondary)}}
  & = T_{02}(\theta_1\rho_2/\rho_1,\rho_1/\theta_1b_g;C_A)
    - T_{02}(\theta_1\rho_2/\rho_1,\zetacut\theta_1;C_A),\\
  R_{\text{mMDT}}(\rho)
  & = T_{02}(\rho,b_i;C_R)-T_{0,2}(\rho,\zetacut;C_R),\\
  R_{\text{SD}}(\rho)
  & = T_{02}(\rho,b_i;C_R)-T_{-\beta,2}(\rho,\zetacut;C_R),
\end{align}
with $b_i=\exp(B_i)$ corresponding to the hard-collinear splittings.

\section{Performance at lower energy}\label{sec:perf-low-pt}

\begin{figure}
  \includegraphics[width=0.48\textwidth,page=3]{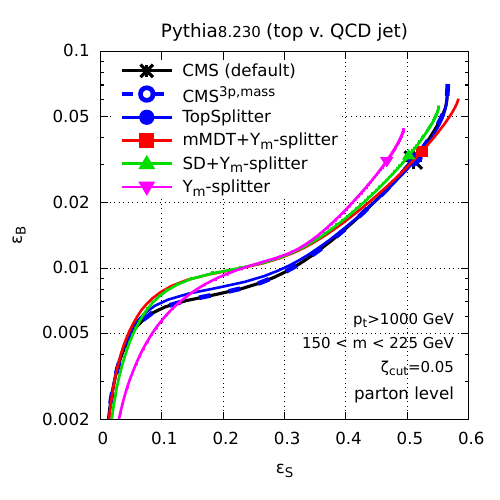}
  \hfill
  \includegraphics[width=0.48\textwidth,page=2]{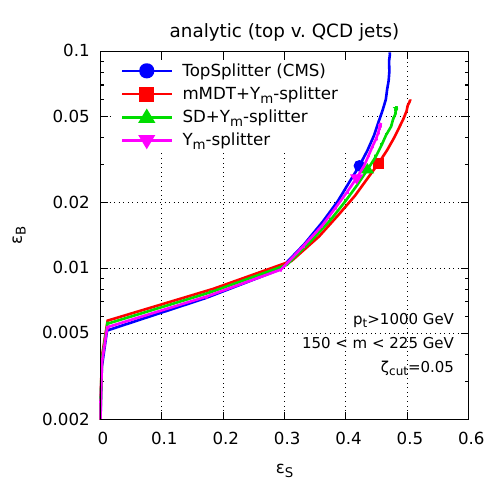}
  \caption{ROC curves obtained for 1~TeV top v. QCD jets when varying $m_{\text{min}}$.
    For the analytic calculation we have assumed a quark fraction of
    $2/3$, roughly corresponding to the matrix elements used in the
    Pythia simulation.  
    The rest is as in Fig.~\ref{fig:rocs}.}\label{fig:rocs-1tev}
\end{figure}

Throughout this paper, for the purpose of verifying our analytical
calculations we have focused on ultra boosted jets with $p_T \sim 2$
TeV. It is therefore natural to check whether our main conclusions
remain valid at lower jet $p_t$.

We show our findings for 1~TeV jets in Fig.~\ref{fig:rocs-1tev}. A
first observation is that due to the somewhat reduced importance of
Sudakov effects, the differences between taggers are less visible than
for the 2 TeV case both in the analytics and in the parton shower
results, which both show a smaller spread between results with
different tagging methods.

As before, the ordering between the performance of the \Ym-splitter
taggers is well reproduced. Differences due to the (pre-)grooming
procedure are also reduced compared to what was seen in
Fig.~\ref{fig:rocs} for 2~TeV jets, which is expected as the
phase-space removed by the grooming procedure is reduced.
The differences between the CMS-related and \Ym-splitter taggers are
not very well reproduced. In the region relevant for phenomenology
this is driven by the efficiency for signal (top) jets, which is
controlled less well in the analytical calculations than for the QCD
background case.
Except at small signal efficiencies where the \TopSplitter performs
marginally worse than the CMS and $\mathrm{CMS^{3p,mass}}$ taggers,
all three taggers perform equivalently at larger signal efficiency
i.e.\ in the phenomenologically relevant region.

Finally, if we go down to yet smaller $p_t$, e.g.\ 500~GeV, the
differences between the taggers are even further suppressed, but our
main conclusion that \TopSplitter is a good overall default choice,
remains unchanged.


\bibliography{bibtexfile} 
\bibliographystyle{JHEP}

\end{document}